\documentclass[reprint,
superscriptaddress,amsmath,amssymb,aps, longbibliography
]{revtex4-2}

\usepackage{graphicx}
\usepackage{dcolumn}
\usepackage{bm}
\usepackage{siunitx}
\usepackage[noend]{algorithmic}
\usepackage{algorithm}
\usepackage[skins]{tcolorbox}
\usepackage{setspace}
\usepackage{blkarray}

\usepackage{xcolor}
\usepackage{soul}
\usepackage{systeme}
\usepackage{footmisc}
\usepackage{mhchem}

\usepackage{scalerel} 

\makeatletter
\def\maketag@@@#1{\hbox{\m@th\normalfont\normalsize#1}}
\makeatother

\newcommand{\blue}[1]{\textcolor{black}{#1}}
\newcommand{\red}[1]{\textcolor{black}{#1}}

\usepackage{booktabs}
\usepackage{hyperref}
\usepackage[capitalize]{cleveref}
\usepackage[utf8]{inputenc}

\usepackage[lining,semibold]{libertine} 
\usepackage{amsthm}
\usepackage[libertine, cmintegrals, bigdelims, vvarbb]{newtxmath}

\usepackage{amsmath}
\usepackage{amsfonts}
\usepackage{mathrsfs}
\usepackage{gensymb}
\usepackage{bbm}
\usepackage{dsfont}

\usepackage{pgfplots}
\usepgfplotslibrary{colormaps}

\usepackage{kbordermatrix}
\renewcommand{\kbldelim}{(}
\renewcommand{\kbrdelim}{)}

\usepackage{chemformula}
\usepackage[caption=false]{subfig}

\usepackage{soul}
\usepackage{xcolor}

\theoremstyle{definition}

\usepackage{scalerel} 

\definecolor{webgreen}{rgb}{0,.5,0}
\definecolor{webbrown}{rgb}{.6,0,0}
\definecolor{grigio}{rgb}{.85,.85,.85} 
\definecolor{RoyalBlue}{rgb}{0.0, 0.14, 0.4}
\definecolor{skyblue1}{rgb}{0.45,0.62,0.81}
\definecolor{skyblue2}{rgb}{0.2,0.39,0.64}
\definecolor{skyblue3}{rgb}{0.13,0.29,0.53}
\definecolor{scarlet1}{rgb}{0.93,0.16,0.16}
\definecolor{scarlet2}{rgb}{0.8,0,0}
\definecolor{scarlet3}{rgb}{0.64,0,0}

\definecolor{g}{gray}{0.50}

\hypersetup{%
    colorlinks=true, linktocpage=true, pdfstartpage=1, pdfstartview=FitV,%
    breaklinks=true, pdfpagemode=UseNone, pageanchor=true, pdfpagemode=UseOutlines,%
    plainpages=false, bookmarksnumbered, bookmarksopen=true, bookmarksopenlevel=1,%
    hypertexnames=true, pdfhighlight=/O,
    urlcolor=webbrown, linkcolor=RoyalBlue, citecolor=webgreen, 
    pdftitle={},%
    pdfauthor={Timur Aslyamov},%
    pdfsubject={},%
    pdfkeywords={},%
    pdfcreator={pdfLaTeX},%
    pdfproducer={LaTeX REVTeX}%
}

\DeclareFontFamily{U}{BOONDOX-calo}{\skewchar\font=45 }
\DeclareFontShape{U}{BOONDOX-calo}{m}{n}{
  <-> s*[1.05] BOONDOX-r-calo}{}
\DeclareFontShape{U}{BOONDOX-calo}{b}{n}{
  <-> s*[1.05] BOONDOX-b-calo}{}
\DeclareMathAlphabet{\mathcalboondox}{U}{BOONDOX-calo}{m}{n}
\SetMathAlphabet{\mathcalboondox}{bold}{U}{BOONDOX-calo}{b}{n}
\DeclareMathAlphabet{\mathbcalboondox}{U}{BOONDOX-calo}{b}{n}

\begin{document}
\title{\blue{Integrated covariances as excess observables weighted by currents and activities}}

\author{Timur Aslyamov}
\email{timur.aslyamov@uni.lu}
\affiliation{Complex Systems and Statistical Mechanics, Department of Physics and Materials Science, University of Luxembourg, 30 Avenue des Hauts-Fourneaux, L-4362 Esch-sur-Alzette, Luxembourg}

\author{Massimiliano Esposito}
\email{massimiliano.esposito@uni.lu}
\affiliation{Complex Systems and Statistical Mechanics, Department of Physics and Materials Science, University of Luxembourg, 30 Avenue des Hauts-Fourneaux, L-4362 Esch-sur-Alzette, Luxembourg}

\date{\today}

\begin{abstract}
Near equilibrium, the symmetric part of the time-integrated steady-state covariance, i.e., the time integral of correlation functions, is governed by the fluctuation-dissipation theorem, while the antisymmetric part vanishes due to Onsager reciprocity.
Far from equilibrium, where these principles no longer apply, we develop a unified formalism for both symmetric and antisymmetric components of integrated covariances. We derive exact, computationally tractable expressions for these quantities, valid in arbitrary nonequilibrium steady states of Markov jump processes and \red{Fokker--Planck equation}.
Both components are expressed in terms of excess observables, a notion central to both statistical physics and reinforcement learning. Furthermore, we establish thermodynamic upper bounds for antisymmetric covariances in terms of (pseudo-)entropy production and cycle affinities.
Finally, we show that the speed up of  self-averaging induced by nonequilibrium drivings which preserve kinetics (activity) is bounded by the cycle affinities (thermodynamic forces). 
\end{abstract}
\maketitle

\section{Introduction}

A hallmark of nonequilibrium steady states (NESS) is the breakdown of the fluctuation-dissipation theorems and Onsager reciprocity \cite{kubo1966fluctuation, kubo2012statistical, stratonovich2012nonlinear, marconi2008fluctuation, forastiere2022linear}. Nonetheless, recent advances have shown that NESS obey a variety of universal relations governing fluctuations, dissipation, and responses to external perturbations. 
These include fluctuation theorems \cite{jarzynski1997nonequilibrium, crooks1999entropy, esposito2009nonequilibrium, seifert2012stochastic}, generalized forms of the fluctuation-dissipation theorem \cite{agarwal1972fluctuation,baiesi2009fluctuations,prost2009generalized,seifert2010fluctuation, baiesi2013update,altaner2016fluctuation,maes2020response,chun2023trade, shiraishi2023introduction, gao2024thermodynamic,tesser2024out,zheng2025nonlinear}, and various expressions for static responses~\cite{cho2000markov,lucarini2016response, santos2020response, falasco2019negative,  mallory2020kinetic, owen2020universal, owen2023size, gabriela2023topologically, aslyamov2024nonequilibrium, aslyamov2024general, harunari2024mutual, cengio2025mutual, khodabandehlou2025affine, floyd2024learning, frezzato2024steady, floyd2024limits, gao2022thermodynamic, bao2024nonequilibrium,  harvey2023universal, ptaszynski2024critical, zheng2025spatial, mitrophanov2025markov,katayama2025diagrammatic}.

In a NESS, physical observables $x(t)$ and $y(t)$ fluctuate around their stationary mean values $\langle x \rangle$ and $\langle y \rangle$. These fluctuations can be characterized by the two-point correlation function (covariance) calculated from the time series of $\Delta x(t) \equiv x(t) - \langle x \rangle$ and $\Delta y(t) \equiv y(t) - \langle y \rangle$: 
\begin{align}
\label{eq:corr}
C_{yx}\blue{(\tau)} \equiv \langle \Delta y(t+\tau) \Delta x(t) \rangle = \langle \Delta y(\tau) \Delta x(0) \rangle\,.
\end{align}
These correlations are independent of $t$ and only depend on the observation time delay $\tau$.
\blue{
They can be decomposed into symmetric (S) and antisymmetric (A) components 
\begin{subequations}
\label{eq:sym-decompose}
\begin{align}
C_{yx}^{S}(\tau) &\equiv C_{yx}(\tau)/2 + C_{xy}(\tau)/2 
= C_{xy}^{S}(\tau) \;,\\ 
C_{yx}^{A}(\tau) &\equiv C_{yx}(\tau)/2 - C_{xy}(\tau)/2  
=-C_{xy}^{A}(\tau) \;.
\end{align}
\end{subequations}
}
The respective symmetrical integrated covariance (SICov) and antisymmetric integrated covariance (AICov) are defined as
\begin{subequations}
\label{eq:Cov-def}
\begin{align}
\label{eq:SICov-def}
\langle\!\langle y, x \rangle\!\rangle_+ &\equiv 
\blue{2\int_0^\infty d\tau C_{yx}^S(\tau)} \;,\\
\label{eq:AICov-def}
\langle\!\langle y, x \rangle\!\rangle_- &\equiv 
\blue{2\int_{0}^\infty d\tau C^A_{yx}(\tau)} \,.
\end{align}
\end{subequations}
These integrated covariances characterize not only the amplitudes of the symmetric and antisymmetric fluctuations but also the system’s overall correlation time, i.e., how long fluctuations persist.
Their sum reads \( \tfrac{1}{2}\big(\langle\!\langle y,x\rangle\!\rangle_{+} + \langle\!\langle y,x\rangle\!\rangle_{-}\big) = \int_{0}^{\infty} dt\, C_{yx}(t) \) and thus determines the static response of \(\langle y\rangle\) and satisfies nonequilibrium fluctuation–dissipation theorems~\cite{agarwal1972fluctuation,baiesi2009fluctuations,prost2009generalized,seifert2010fluctuation,baiesi2013update,maes2020response,zheng2025nonlinear}; see \cref{sec:response}. 

The SICov \blue{coincides with the zero-frequency power spectral density, a central measure of how fluctuations are distributed across timescales~\cite{landi2023current,aslyamov2026macroscopic}}. It has been shown to play an important role in various thermodynamic bounds and in inferring the statistical properties of physical observables that are difficult to measure directly~\cite{barato2015thermodynamic, gingrich2016dissipation, pietzonka2016universal, pietzonka2017finite, horowitz2017proof, falasco2020unifying, horowitz2020thermodynamic, vu2020entropy, van2023thermodynamic, di2018kinetic, van2022unified, shiraishi2021optimal, dechant2019arxiv, dechant2020fluctuation, ptaszynski2024dissipation, kwon2024fluctuation}. 
The recently discovered fluctuation-response relations provide an exact expression for the SICov in terms of static responses~\cite{aslyamov2025frr, ptaszynski2024frr, ptaszynski2025frr-mixed}, which, close to equilibrium, reduces to the fluctuation-dissipation theorem \cite{kubo1966fluctuation,kubo2012statistical}. 

The AICov constitutes a measure of the degree of nonreciprocity in NESS, since at equilibrium, Onsager reciprocity implies that $C^A_{xy}(\tau)=0$ and the AICov vanishes. 
Using $C^A_{xy}(\tau)$ as a measure of the degree of nonequilibrium is an old idea \cite{tomita1974irreversible, steinberg1986time, qian2004fluorescence, battle2016broken}. It has more recently been studied in classical \cite{liang2023thermodynamic} and quantum \cite{van2024dissipation} systems. In its short lag time limit $\tau \to 0$, it has been shown to be bounded by the cycle affinity \cite{ohga2023thermodynamic}, by entropy production \cite{shiraishi2023entropy}, and by the system activity \cite{gu2024thermodynamic}. 
\blue{Moreover, for the limit $\tau\to 0$, the antisymmetric $C_{yx}^A(\tau)$ relates to odd diffusion  \cite{yasuda2022time}.  
}
To our knowledge, no insightful bounds or exact relations are known for its time-integrated expression, namely the AICov.

In this Paper, within the frameworks of Markov jump processes and \red{Fokker--Planck equation}, we present a formalism that unifies the study of the SICov [\cref{eq:SICov-def}] and the AICov [\cref{eq:AICov-def}]. We use it to derive exact expressions for the SICov and AICov in terms of \textit{excess observables} associated to $x$ and $y$. These excess observables originate from works in stochastic thermodynamics \cite{komatsu2008steady, komatsu2009representation} and are known as ``bias'' in the context of reinforcement learning \cite{ortner2024note}. \red{In the context of Markov jump processes}, they have recently been used in studies on nonequilibrium heat capacities \cite{khodabandehlou2023nernst,khodabandehlou2024close,bogers2025negative}, and mathematical results \cite{khodabandehlou2024poisson, ortner2024note} have revealed their relation to mean first passage times (MFPTs)~\cite{redner2001guide}. 
\red{
We derive SICov/AIcov within Fokker--Planck formulation by taking the continuum limit with diffusive scaling of the Markov jump processes.  
}
We also use our exact expressions to derive thermodynamic bounds for the AICov in terms of (pseudo-)entropy production and cycle affinity. 
As an application of our theory, we study the speed up of self-averaging in nonequilibrium systems~\cite{neal2004improving,suwa2010markov,sun2010improving,duncan2016variance,coghi2021role,dechant2023thermodynamic} and provide a physical interpretation of the nonreversible Markov Chain Monte Carlo (MCMC) algorithm~\cite{bierkens2016non}.

\section{Descrete space: Markov jump processes}

\red{We derive exact expressions and thermodynamic bounds for SICov and AICov for Markov jump processes.} 

\subsection{Setup}

\blue{We consider a Markov jump process over $N$ discrete states labeled by $n \in \{1,\dots, N\}$. 
The trajectory of the system between time $t=0$ and $t=T$ is denoted $n_t$.
We introduce an observable $x$ that takes the values $x_n$ on each state $n$ and defines the vector $\boldsymbol{x}=(x_1,\dots,x_N)^\intercal$. 
The value of the observable along a trajectory is $x(t)\equiv x_{n_t}$.
For a cell receptor in different signaling states $n$, $x(t) = \sum_{n} \delta_{nn_t}$ measures the activation of the receptors~\cite{berg1977physics,lang2014thermodynamics,harvey2023universal}. For a walker on a 1D lattice with step size $dx$, 
$x(t)=n_t dx$ is the spatial coordinate. 
}
\blue{
The probability distribution of the states, $\boldsymbol{P}=(\dots,P_n,\dots)^\intercal$, satisfies to the master equation $d_t \boldsymbol{P}(t) = \mathbb{W}\cdot\boldsymbol{P}(t)$ with formal solution
\begin{align}
\label{eq:sol-formal}
    \boldsymbol{P}(t) = e^{\mathbb{W}t}\cdot\boldsymbol{P}(0)\,,
\end{align}
where $\boldsymbol{P}(0)$ is the initial distribution and where the off-diagonal elements $W_{mn}$ denotes the transition rate from state $n$ to state $m$, and the diagonal  elements are $W_{nn}=-\sum_{m}W_{mn}$.
}
We assume that the Markov jump process is ergodic and always relaxes to the unique steady-state probability $\boldsymbol{P}^\text{ss}\equiv \boldsymbol{P}(\infty)$ satisfying $\sum_{n}W_{mn}P^\text{ss}_n = 0$. 
Therefore, the ensemble average of an observable $x(t)$ coincides with its time average along any stochastic trajectory, regardless of its initial condition as
\begin{align}
\label{eq:mean}
    \langle x \rangle  \equiv \sum_n x_n P^\text{ss}_n &= \lim_{T \to \infty}\frac{1}{T}\int_0^T dt\,\blue{x(t)} \,,
\end{align}
\blue{and is directly accessible experimentally. 
}
\blue{
For examples, $\langle x \rangle$ could be the fraction of total measurement time $T$ during which the receptor is signaling or the time-averaged position of the random walker.
}

\subsection{Excess observables}

\subsubsection{Definition}

\blue{We now consider an ensemble of trajectories conditioned on the initial condition $n_0 = m$, implying $x(0)=x_{m}$. The average of $x(t)$ over this conditional ensemble reads
\begin{align}
\label{eq:mean-cond}
    \langle x(t) |m \rangle &\equiv 
    \sum_{n} x_n P_n(t|m)\,,
\end{align}
where $P_n(t|m) \equiv [e^{t\mathbb{W}}]_{nm}$ is the probability to start in $m$ and end in $n$ after a time $t$. 
The difference between the conditional and regular average reads 
\begin{align}
\label{eq:means-CLimit}
    \langle x(t) |m \rangle - \langle x \rangle = \big[\boldsymbol{x} \cdot (e^{\mathbb{W}t} - \boldsymbol{P}^\text{ss}\boldsymbol{1}^\intercal)\big]_{m}\,,
\end{align}
which tends to zero over long time since the process is ergodic and the memory of the initial condition gets lost: $\lim_{t \to \infty} P_n(t|m) = P_n^\text{ss}$, implying  $\lim_{t \to \infty} \langle x (t) |m \rangle = \langle x \rangle$. 
This quantity is also zero after stationary averaging over all initial conditions since such an average restores the unconditioned ensemble average from \cref{eq:mean}: $\sum_m \langle x(t)|m \rangle P^\text{ss}_m = \langle x \rangle$.
}

\blue{The time integral of \cref{eq:means-CLimit}, coined \textit{excess observable}, is thus a quantity of interest capturing the intensity of the dynamical self-averaging of the observable $x$ as
}
\begin{align}
\label{eq:X-def}
    X_{m} \equiv \int_0^\infty dt \big(\blue{\langle x(t) |m \rangle} - \langle x \rangle \big)\,.
\end{align}
Graphically, it measures the shaded area in \cref{fig:fig-1}(a). It naturally vanishes after stationary averaging: $\langle X \rangle  \equiv \sum_m X_m P^\text{ss}_m = 0$. 
Excess observables have been related to notions of excess heat and entropy \cite{komatsu2008steady, komatsu2009representation, khodabandehlou2023nernst, bogers2025negative, khodabandehlou2024close}. They also appear in reinforcement learning as ``bias'' \cite{ortner2024note}: the reward $r_n$ is a state observable with an average $\langle r \rangle$ that is independent of the initial state, and the bias $R_m$ quantifies the transient advantage of starting in a particular state $m$. 
\blue{Recently they have also been related to the mean and variance of transition times \cite{garilli2025interrelation}.
Operationally, excess observables \eqref{eq:X-def} can be measured by preparing the system in a prescribed initial state and tracking the deviation between the time-averaged observable from its steady-state mean,
for instance, single-molecule experiments where a molecular motor is initialized in a given chemical or conformational state; 
receptor–ligand binding kinetics where the receptor is prepared in a specific signaling state; 
colloidal particles in optical traps where the initial particle position or orientation is controlled. 
}

\subsubsection{Linking excess observables with Drazin inverse}

Using \cref{eq:X-def,eq:means-CLimit}, excess observables $\boldsymbol{X} \equiv (X_1,\dots,X_N)^\intercal$ can be expressed as
\begin{align}
\label{eq:X-Drazin}
    \boldsymbol{X}^\intercal = \boldsymbol{x}\cdot \int_0^\infty dt  e^{\mathbb{W}t}\cdot\big[\mathbb{1}  -\boldsymbol{P}^\text{ss}\boldsymbol{1}^\intercal \big] = - \boldsymbol{x}\cdot \mathbb{W}^D\,,
\end{align}
in terms of the Drazin inverse of the rate matrix \cite{landi2023current,crook2018drazin}
\begin{subequations}
\begin{align}
\label{eq:Drazin-def}
    \mathbb{W}^D &\equiv - \int_0^\infty e^{\tau \mathbb{W}}\cdot\big[\mathbb{1}  -\boldsymbol{P}^\text{ss}\boldsymbol{1}^\intercal \big]d\tau\,,\\
\label{eq:Drazin-property}
    \mathbb{W}^D \cdot\mathbb{W} &= \mathbb{W}\cdot\mathbb{W}^D = \mathbb{1} - \boldsymbol{P}^\text{ss}\boldsymbol{1}^\intercal\,,
\end{align}
\end{subequations}
where $\mathbb{1}$ is the identity matrix. 
\blue{Since $\mathbb{W}^D$ has the same left $\boldsymbol{\ell}^{(n)}$ and right eigenvectors $\boldsymbol{r}^{(n)}$ as $\mathbb{W}$, and its eigenvalues are $\{0, \lambda_2^{-1}, \dots, \lambda_N^{-1}\}$, where $\lambda_{n>1}$ are nonzero eigenvalues of $\mathbb{W}$, \cref{eq:X-Drazin} can be rewritten in terms of the relaxation time scales $\text{Re}~\lambda_n<0$ and the characteristic oscillation frequencies $1/\text{Im}~\lambda_n$ of the system as $X_m = -\sum_{n>1}\boldsymbol{x} \cdot\boldsymbol{r}^{(n)}\ell_m^{(n)}/\lambda_n$.}  

\subsubsection{Linking excess observables with MFPTs}

Here, for self-consistency, we re-derive recent mathematical results \cite{khodabandehlou2024poisson,ortner2024note}, stating that excess observables can be expressed, up to a constant independent of the initial state $m$, as
\begin{align}
\label{eq:X-MFTS}
     X_{m} = - \sum_{n}x_nP^\text{ss}_n T_{nm} + \text{const.} 
     \,,
\end{align}
where 
$T_{nm} \equiv \int  t_{nm} f(t_{nm}) dt_{nm}$ is the MFPT from state $m$ to state $n$ with
$t_{nm} \equiv \inf_{t}(n_t = n|n_0 = m)$ being the first passage time and $f(t_{nm})$ its distribution function~\cite{redner2001guide}.

\textit{Proof---}The idea is to show that both the expressions for excess observables, \cref{eq:X-def} and \cref{eq:X-MFTS}, satisfy the Poisson equation 
\begin{align}
\label{eq:Poisson}
    \mathbb{W}^\intercal \boldsymbol{X} + \Delta\boldsymbol{x} = 0\,.
\end{align}

Since we have shown in the main text that \cref{eq:X-def} is equivalent to \cref{eq:X-Drazin} and since
\begin{align}
    \mathbb{W}^\intercal \boldsymbol{X} = -\mathbb{W}^\intercal(\mathbb{W}^D)^\intercal \Delta\boldsymbol{x} = -(\mathbb{1} - \boldsymbol{1}(\boldsymbol{P}^\text{ss})^\intercal)\Delta\boldsymbol{x}=-\Delta\boldsymbol{x}\,,
\end{align}
where we used \cref{eq:Drazin-property} and $(\boldsymbol{P}^\text{ss})^\intercal\Delta\boldsymbol{x} = 0$, we just proved that \cref{eq:X-def} satisfies \cref{eq:Poisson}.

We now turn to \cref{eq:X-MFTS}. By inserting it in the left-hand side of \cref{eq:Poisson} we get
\begin{align}
\label{eq:Poisson-MFPT-1}
    \sum_m W_{m n } X_m + \Delta x_n =  -\sum_k x_k P^\text{ss}_k \sum_m W_{m n }T_{k m} +
    \Delta x_n\,.
\end{align}
We then recall the expression of MFPTs in terms of the Drazin inverse \cite{harvey2023universal,khodabandehlou2025affine,bao2024nonequilibrium} 
\begin{align}
\label{eq:MFPT-Drazin}
    T_{km} = \frac{W_{km}^D-W_{kk}^D}{P^\text{ss}_k}\,,
\end{align}
which implies that
\begin{align}
\label{eq:MFPT-relation}
    \sum_{m}T_{km} W_{mn} &= \frac{1}{P^\text{ss}_k}\big[(\delta_{kn}-P^\text{ss}_k) - W_{kk}^D\sum_m W_{mn}\big]
    =\frac{\delta_{kn}}{P^\text{ss}_k} - 1\,,
\end{align}
where we used $\sum_m W_{km}^DW_{mn}=(\mathbb{1}-\boldsymbol{P}^\text{ss}\boldsymbol{1}^\intercal)_{kn}$ and $\sum_m W_{mn} = 0$. 
Combining \cref{eq:MFPT-relation} with \cref{eq:Poisson-MFPT-1}, we arrive at
\begin{align}
    \sum_m W_{m n } X_m + \Delta x_n = - \sum_k x_k (\delta_{kn} - P^\text{ss}_k) + \Delta x_n = 0\,,
\end{align}
which proves that \cref{eq:X-MFTS} solves \cref{eq:Poisson}.

\subsection{Linking integrated covariances and excess observables to static response}
\label{sec:response}

The static response describes the change of steady state probability due to the perturbations of a model parameter $\varepsilon$ controlling the rates $\mathbb{W}(\varepsilon)$. Following \cite{cho2000markov, harvey2023universal, ptaszynski2024critical, ptaszynski2024frr, bao2024nonequilibrium}, it can be written as
\begin{align}
\label{eq:response-1}
    \frac{d\boldsymbol{P}^\text{ss}}{d\varepsilon} = - \mathbb{W}^D \frac{\partial \mathbb{W}}{\partial \varepsilon}\boldsymbol{P}^\text{ss}\,,
\end{align}
which is an alternative approach to \cite{aslyamov2024nonequilibrium,aslyamov2024general}. 

\blue{We first use \cref{eq:response-1} to recover steady-state fluctuation--dissipation relations, including Agarwal’s result~\cite{agarwal1972fluctuation} and the Seifert--Speck formulation~\cite{seifert2010fluctuation}. Using \cref{eq:Drazin-def} with $\boldsymbol{1}^\intercal \partial_\varepsilon\mathbb{W}=0$, we rewrite \cref{eq:response-1} as
\begin{align}
\label{eq:response-cov}
\frac{d\langle y\rangle}{d\varepsilon}
&= \int_0^\infty dt\, C_{yx}(t) \nonumber\\
&= \tfrac12\!\left[\langle\!\langle y,x\rangle\!\rangle_+ + \langle\!\langle y,x\rangle\!\rangle_-\right], 
\end{align}
where $\boldsymbol{x}\equiv\mathbb{P}^{-1}\partial_\varepsilon\mathbb{W}\,\boldsymbol{P}^\text{ss}$ is the Agarwal observable \cite{agarwal1972fluctuation}. To connect with the stochastic-entropy observable of~\cite{seifert2010fluctuation}, we perturb the rate matrix $\mathbb{W}(\varepsilon)$ while keeping the initial state at $t=0$ equal to the unperturbed NESS, $\boldsymbol{P}(0)\equiv\boldsymbol{P}^\text{ss}$. After switching on the perturbation, we have $\boldsymbol{P}(t)=e^{\mathbb{W}(\varepsilon)t}\boldsymbol{P}^\text{ss}$. Defining $s_n(t)=-\log P_n(t)$, one finds 
\begin{align}
\partial_\varepsilon\partial_t s_n\big|_{t=0}
= -[\mathbb{P}\,\partial_\varepsilon\mathbb{W}\,\boldsymbol{P}^\text{ss}]_n\,,
\end{align}
implying that $x_n=-\partial_\varepsilon\dot{s}_n(0)$ transforms \cref{eq:response-cov} to fluctuation-disspation relation of Seifert and Speck~\cite{seifert2010fluctuation}.}

Second, we express the difference $X_k - X_l$ to static responses.
Choosing $\varepsilon\in \{W_{mn},W_{nm}\}$, $\partial_\varepsilon \mathbb{W}$ reads 
\begin{align}
\label{eq:response-2}
\partial_{W_{mn}}\mathbb{W}=
\begin{blockarray}{ccccccccc}
& &  \color{gray} \dots & \color{gray} n & \color{gray}\dots & \\
\begin{block}{cc (ccccccc)}
\color{gray} \vdots & &  \phantom{0} & \phantom{0} & \phantom{0} & \phantom{0} \\
\color{gray} m & & \phantom{0} & 1 & \phantom{0} & \phantom{0} \\
\color{gray} \vdots & &  \phantom{0} & \phantom{0} & \phantom{0} & \phantom{0} \\
\color{gray} n & & \phantom{0} & -1 & \phantom{0} & \phantom{0} \\
\color{gray} \vdots & &  \phantom{0} & \phantom{0} & \phantom{0} & \phantom{0}\\
\end{block}
\end{blockarray}\,\,,
\quad
\partial_{W_{nm}}\mathbb{W}=
\begin{blockarray}{ccccccccc}
& &  \color{gray} \dots & \color{gray} m & \color{gray}\dots & \\
\begin{block}{cc (ccccccc)}
\color{gray} \vdots & &  \phantom{0} & \phantom{0} & \phantom{0} & \phantom{0} \\
\color{gray} m & & \phantom{0} & -1 & \phantom{0} & \phantom{0} \\
\color{gray} \vdots & &  \phantom{0} & \phantom{0} & \phantom{0} & \phantom{0} \\
\color{gray} n & & \phantom{0} & 1 & \phantom{0} & \phantom{0} \\
\color{gray} \vdots & &  \phantom{0} & \phantom{0} & \phantom{0} & \phantom{0}\\
\end{block}
\end{blockarray}\,\,,
\end{align}
where only non-zero elements are shown. Combining \cref{eq:response-1,eq:response-2}, we find
\begin{align}
\label{eq:response-3}
    \frac{dP^\text{ss}_k}{dW_{mn}} = -(W^D_{km} - W^D_{kn})P^\text{ss}_n\,,\,\,
    \frac{dP^\text{ss}_k}{dW_{nm}} = (W^D_{km} - W^D_{kn})P^\text{ss}_m \;.
\end{align}

We proceed using the Arrhenius-like parameterization 
\begin{align}
\label{eq:response-4}
    W_{mn}=e^{B_{mn}+S_{mn}/2}\,,\quad W_{nm}=e^{B_{mn} - S_{mn}/2}\,,
\end{align}
where $B_{mn}=B_{nm}$ and $S_{mn}=-S_{nm}$ are the symmetric and antisymmetric parts of the rates. The physical interpretation of these quantities is discussed in \cite{aslyamov2025frr,ptaszynski2024frr}. Then, using \cref{eq:response-1,eq:response-3}, the symmetric response can be written as
\begin{align}
\label{eq:response-5}
    \frac{dP^\text{ss}_k}{dB_{mn}}&=W_{mn}\frac{dP^\text{ss}_k}{dW_{mn}} + W_{nm}\frac{dP^\text{ss}_k}{dW_{nm}} \nonumber\\
    &= -J_{mn}(W^D_{km} - W_{kn}^D) = -J_{mn}P^\text{ss}_k(T_{km} - T_{kn})\,,
\end{align}
where \cref{eq:MFPT-Drazin} was used for the last identity. 
Multiplying both sides of \cref{eq:response-5} by $x_k$, and calculating the sum over $k$, we arrive at
\begin{align}
    \label{eq:response-6}
    \sum_{k}x_k\frac{dP^\text{ss}_k}{dB_{mn}} = J_{mn}(X_m - X_n)\,.
\end{align}
We notice that a similar expression can be found for antisymmetric perturbation $S_{mn}$

\begin{figure}
    \centering
    \includegraphics[width=\linewidth]{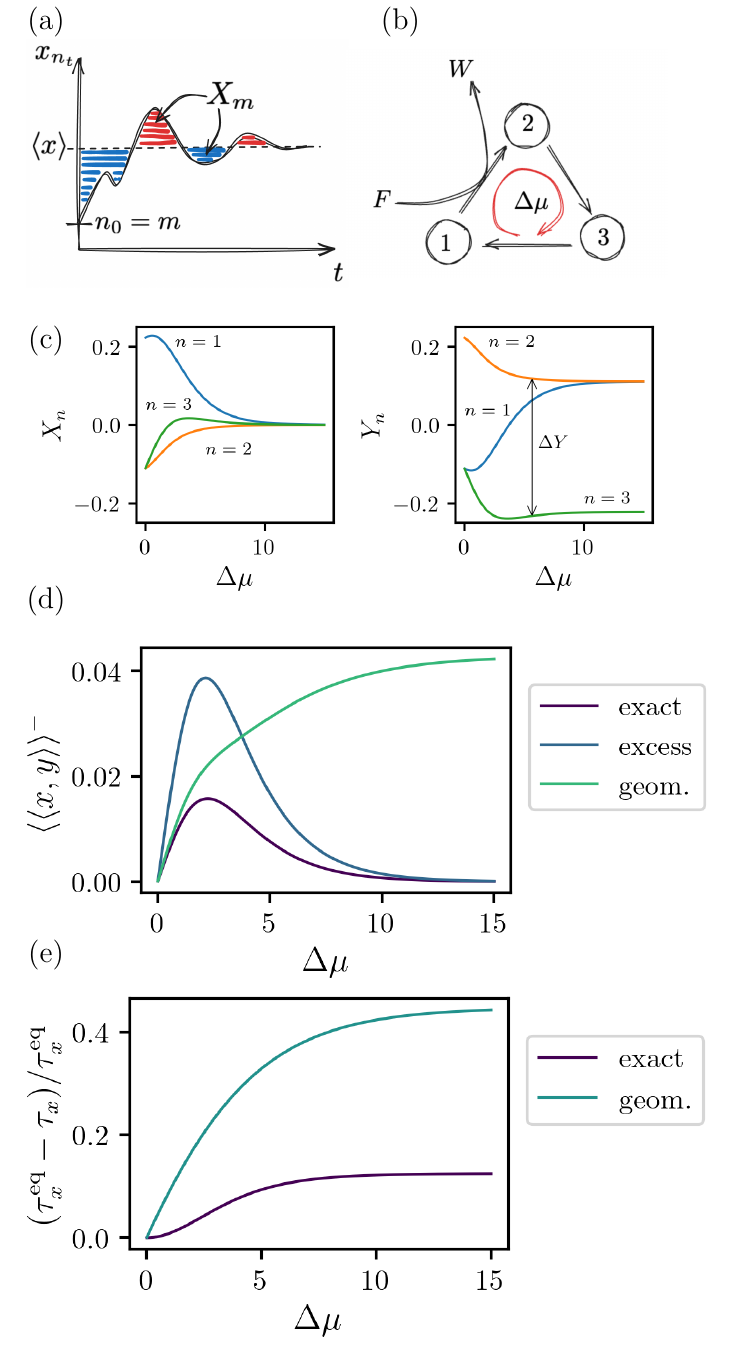}
    \caption{
    (a): Sketch of a generic excess observable showing that its value $X_n$ corresponds to the net area of the highlighted region (the area above is positive and the one below negative).
    (b): Molecular motor with three states, $n\in\{1,2,3\}$, driven by fuel $F$ and waste $W$ generating a nonequilibrium drive $\Delta\mu$ on the transition from $1$ to $2$. 
    (c): Excess observables $\boldsymbol{X}=(X_1,X_2,X_3)^\intercal$ and $\boldsymbol{Y} = (Y_1,Y_2,Y_3)^\intercal$ associated to observables $x=(1,0,0)^\intercal$ and $y = (0,1,0)^\intercal$ for various chemical potentials $\Delta \mu$.
    (d): AICov (purple curve) [\cref{eq:AICov-excess,eq:AICov-def}], geometric bound (green) [\cref{eq:bound-geometric}] and excess bound (blue) [\cref{eq:bound-excess}].
    (e): Purple curve is the relative speed up $(\tau_x^\text{eq}-\tau_x)/\tau_x^\text{eq}$ as a function of thermodynamic force $\Delta\mu$ calculated from \cref{eq:speed-up-exact}, where $\tau_x^\text{eq}$ corresponds to the equilibrium partner with the same probability and activity. Green curve is the goemetric bound \cref{eq:bound-tau_x} with $\mathcal{M}=\tanh(\Delta\mu/3)/\tan(\pi/6)$.
    Parameters: Matrix $\mathbb{W}$ has off-diagonal elements $W_{21}=e^{\Delta\mu/2}$, $W_{12}=e^{-\Delta\mu/2}$, $W_{31}=W_{13}=W_{23}=W_{32}=1$. 
    }
    \label{fig:fig-1}
\end{figure}

\subsection{Central Result: Exact relations for integrated covariances}
\label{sec:discr-exact-results}
\subsubsection{Symmetric and antisymmetric integrals}

The covariance (\ref{eq:corr}) between two state observables $x_n$ and $y_n$ reads
\begin{align}
\label{eq:corr-Markov}
    C_{yx}(\tau) \equiv \Delta\boldsymbol{y}\cdot e^{\tau \mathbb{W}} \cdot \mathbb{P} \cdot \Delta \boldsymbol{x}\,,
\end{align}
where $\mathbb{P}\equiv \text{diag}(P^\text{ss}_1,\dots,P^\text{ss}_n)$ is the diagonal matrix.
In \cref{sec:derivation-matC} we show that 
the SICov, \cref{eq:SICov-def}, and AICov, \cref{eq:AICov-def}, can be written as
\begin{subequations}
\begin{align}
\label{eq:SICov-matC}
    \langle\!\langle y, x \rangle\!\rangle_+ &= \Delta\boldsymbol{y}\cdot\mathbb{C}^+\cdot\Delta\boldsymbol{x} \,,\\
\label{eq:AICov-matC}
    \langle\!\langle y, x \rangle\!\rangle_- &= \Delta\boldsymbol{y}\cdot\mathbb{C}^-\cdot\Delta\boldsymbol{x} \,,  
\end{align}
\end{subequations}
in terms of the unified matrices
\begin{align}
\label{eq:matC-results-1}
    \mathbb{C}^\pm 
    &=-\mathbb{W}^D\cdot\big[\mathbb{P}\cdot\mathbb{W}^\intercal \pm \mathbb{W}\cdot\mathbb{P}\big]\cdot(\mathbb{W}^D)^\intercal\,.
\end{align}

The activity (also known as traffic or frenesy \cite{maes2020frenesy}) and the current of a transition between a pair of states characterize, respectively, the total and the net number of times a transition occurs per unit time. 
The \textit{activity matrix} $\mathbb{A}$ has off-diagonal elements $A_{kl} \equiv W_{kl}P^\text{ss}_l + W_{lk}P^\text{ss}_k$ and diagonal elements $A_{ll} \equiv -\sum_{k\neq l}A_{kl}$.
The \textit{current matrix} $\mathbb{J}$ has off-diagonal elements $J_{kl} \equiv W_{kl}P^\text{ss}_l  - W_{lk}P^\text{ss}_k$ and zero diagonal elements $J_{ll} \equiv 0$, where $J_{kl}$ is the net probability current from state $l$ to state $k$. 
Activity and current matrices play a central role in stochastic thermodynamics \cite{maes2008canonical, maes2020frenesy, seifert2012stochastic}. 
Remarkably, from \cref{eq:matC-results-1}, we find that the unified matrices determining the SICov and AICov in \cref{eq:SICov-matC,eq:AICov-matC}, are respectively expressed in terms of the activity and current matrices
\begin{subequations}
\begin{align}
\label{eq:matC-plus-result-1}
&\mathbb{C}^+ = - \mathbb{W}^D\cdot\mathbb{A}\cdot(\mathbb{W}^D)^\intercal\,, \\
\label{eq:matC-minus-result-1}
&\mathbb{C}^- = \mathbb{W}^D \cdot\mathbb{J}\cdot(\mathbb{W}^D)^\intercal \;.
\end{align}
\end{subequations}
While \cref{eq:matC-plus-result-1} agrees with the symmetric covariance matrices derived in Refs.~\cite{ptaszynski2024frr, lapolla2020spectral, lapolla2019manifestations, lapolla2018unfolding}, to our knowledge, \cref{eq:matC-minus-result-1} was not previously known. \cref{eq:matC-plus-result-1,eq:matC-minus-result-1} thus constitutes our first main result.

\subsubsection{SICov and AICov in terms of excess observables}

Our second main result is an exact relation for the SICov in terms of excess observables weighted by the activity.
Using \cref{eq:matC-plus-result-1} in \cref{eq:SICov-matC}, we find
\begin{align}
    \label{eq:SICov-excess-1}
    \langle\!\langle y, x \rangle\!\rangle_+ &=- \boldsymbol{Y}\cdot\mathbb{A}\cdot\boldsymbol{X} = -\boldsymbol{X}\cdot\mathbb{A}\cdot\boldsymbol{Y}\,,
\end{align}
where, following \cref{eq:X-Drazin}, $\boldsymbol{X}^\intercal =- \Delta\boldsymbol{x}\cdot \mathbb{W}^D$ and
$\boldsymbol{Y}=- (\mathbb{W}^D)^\intercal \cdot \Delta\boldsymbol{y}$. 
Rewriting $-\sum_{kl}X_{k}A_{kl}Y_l$ with $A_{ll} = -\sum_{k\neq l}A_{kl}$, and reducing the summation to $\sum_{k<l}$, we obtain
\begin{subequations}
\label{eq:SICov-excess-2}
\begin{align}
\langle\!\langle y, x \rangle\!\rangle_+ &= \sum_{k<l}A_{kl}(X_{k} - X_{l})(Y_{k} - Y_{l})\,,\\
\label{eq:var-excess}
\langle\!\langle x, x \rangle\!\rangle_+ &= \sum_{k<l}A_{kl}(X_{k} - X_{l})^2\,,
\end{align}
\end{subequations}
where $\langle\!\langle x, x \rangle\!\rangle_+$ is the variance of the observable $\boldsymbol{x}$. 
Using \cref{eq:response-6}, we express the SICov, \cref{eq:SICov-excess-2}, in terms of static responses as
\begin{align}
    \langle\!\langle y, x \rangle\!\rangle_+ &= \sum_{n,m}x_ny_m\sum_{k<l}\frac{A_{kl}}{J_{kl}^2}\frac{dP^\text{ss}_n}{dB_{kl}}\frac{dP^\text{ss}_m}{dB_{kl}}\,,
\end{align}
which is reminiscent of the recently derived fluctuation response relations \cite{aslyamov2025frr, ptaszynski2024frr, ptaszynski2025frr-mixed}.

Our third main result is an exact expression for the AICov in terms of excess observables. 
Using \cref{eq:matC-minus-result-1,eq:X-Drazin}, we express the AICov in \cref{eq:AICov-matC} as
\begin{align}
\label{eq:AICov-excess}
    \langle\!\langle y, x \rangle\!\rangle_- &= \boldsymbol{Y} \cdot \mathbb{J} \cdot \boldsymbol{X} = - \boldsymbol{X}\cdot \mathbb{J}\cdot\boldsymbol{Y}\,.
\end{align}
In coordinate form, $\langle\!\langle y, x \rangle\!\rangle_- = \sum_{kl} Y_k X_l J_{kl}$ quantifies the AICov through weighted currents between states. 
At equilibrium, since $J_{kl}=0$ for $\forall k,l$, \cref{eq:AICov-excess} implies Onsager reciprocity $\langle\!\langle y, x \rangle\!\rangle_- = 0$. 
Away from equilibrium, since $J_{kl} = - J_{lk}$, we obtain
\begin{align}
\label{eq:AICov-excess-2}
    \langle\!\langle y, x \rangle\!\rangle_- = \sum_{k<l} (Y_k X_l - Y_l X_k) J_{kl} \,.
\end{align}
Doing so reveals that the AICov measures the sum over all pairs of connected states of the nonreciprocities in excess observables, $(Y_k X_l - Y_l X_k)$, weighted by the corresponding current, $J_{kl}$. The AICov can also be expressed in terms MFPTs using \cref{eq:X-MFTS}. Indeed, considering the asymmetry between one-point observables $\boldsymbol{x}=(x_n)$ and $\boldsymbol{y}=(y_m)$, we find that
\begin{align}
    C^-_{nm} &= \pi_n \pi_m \sum_{kl}T_{mk}T_{nl}J_{kl}\nonumber\\
    &=\pi_n \pi_m \sum_{k<l}(T_{mk}T_{nl}-T_{ml}T_{nk})J_{kl}\,,
\end{align}
thus revealing that the AICov is also a measure of non-reciprocity in MFPTs.

\textit{Importance Remark---}Let us comment on the meaning and importance of \cref{eq:SICov-excess-2,eq:AICov-excess-2}.
The AICov and SICov are easily measurable \textit{macroscopic} properties of the system, while the activities $A_{nm}$ and currents $J_{nm}$ are \textit{microscopic} quantities defined at the level of individual transitions. 
\textit{Excess observables} $X_m$ characterize macroscopic correlations but for systems initially prepared in a given microscopic state. They thus connect the microscopic to the macroscopic level in \cref{eq:SICov-excess-2,eq:AICov-excess-2}. 
More specifically, \cref{eq:AICov-excess-2} relates an antisymmetric macroscopic quantity quantifying nonreciprocity, the AICov, to the microscopic currents, antisymmetric quantities which quantify the break of detailed balance (time-reversal symmetry) occurring out-of-equilibrium. 
\cref{eq:SICov-excess-2} instead connect a less familiar symmetric macroscopic quantity, the SICov, to a symmetric microscopic property, the activity of the transitions.

\subsection{Thermodynamic Tradeoffs}
\subsubsection{Geometric bound}

We now make use of the geometric approach to thermodynamic bounds developed in \cite{ohga2023thermodynamic} on \cref{eq:AICov-excess-2,eq:SICov-excess-2}. We consider the ratio
\begin{align}
\label{eq:ratio-1}
    \frac{\langle\!\langle y, x \rangle\!\rangle_- }{\langle\!\langle x, x \rangle\!\rangle_+ + \langle\!\langle y, y \rangle\!\rangle_+} = \frac{2 \sum_{k<l} J_{kl} \Omega_{kl} }{\sum_{k<l}A_{kl} L_{lk}^2}\,,
\end{align}
where $L_{lk} \equiv \sqrt{( X_l -  X_k)^2+( Y_l -  Y_k)^2}$ and $\Omega_{kl} \equiv   \tfrac{1}{2}(X_{l} Y_{k} - X_{k}Y_{l})$. Using Eqs.~(3) and (S7) in \cite{ohga2023thermodynamic} for the right-hand side of \cref{eq:ratio-1}, we find the \textit{geometric bound} as
\begin{align}
\label{eq:bound-geometric}
    |\langle\!\langle y, x \rangle\!\rangle_-| &\leq \frac{\langle\!\langle x, x \rangle\!\rangle_+ + \langle\!\langle y, y \rangle\!\rangle_+}{2}\mathcal{M}\,,\nonumber\\
    \mathcal{M}&\equiv \max_c\frac{\tanh\tfrac{|\mathcal{F}_c|}{2n_c}}{\tan\tfrac{\pi}{n_c}}\,,
\end{align}
where the maximum is taken over all simple (closed paths of states without repetition) 
cycles of the graph associated with the Markov jump process, $\mathcal{F}_c$ is the affinity of the cycle (defined as the natural logarithm of the ratio between the product of the rates along the forward orientation of the cycle and the product of the rates along the backward orientation), and $n_c$ is the number of states in the cycle.
\blue{The cycle affinities directly relate to thermodynamic forces for thermodynamically consistent dynamics \cite{rao2018conservation}. At equilibrium $\mathcal{F}_c \to 0$ implies $\mathcal{M} \to 0$ and thus Onsager reciprocity is recovered from \cref{eq:bound-geometric}.}
We note that \cref{eq:bound-geometric} could be further improved using uniform cycles \cite{pietzonka2016affinity}, as discussed in \cite{ohga2023thermodynamic}. We also note that it is possible to bound the ratio in \cref{eq:ratio-1} using the geometrical approach from \cite{shiraishi2023entropy}.

\subsubsection{Excess bound}
To derive a second inequality, termed the \textit{excess bound}, we first note that $\sum_{k<l}J_{kl}(X_{k}Y_{k} - X_{l}Y_{l})= 0$ and rewrite \cref{eq:AICov-excess-2} as
\begin{align}
\label{eq:bounds-excess-1}
    \langle\!\langle y, x \rangle\!\rangle_-
    &= \sum_{k<l} J_{kl} (X_l - X_k) (Y_{k} + Y_{l})  \nonumber\\
    &= \sum_{k<l} \sqrt{A_{kl}}(X_l - X_k) \frac{J_{kl}}{\sqrt{A_{kl}}}(Y_{k} + Y_{l} - \mu_1 - \mu_2)\nonumber\\
    &\leq \sqrt{\langle\!\langle x, x \rangle\!\rangle_+}
    \big[\sqrt{ V_1} + \sqrt{ V_2}\big]\sqrt{\dot{\Pi}}
    \,,
\end{align}
which is defined up to an arbitrary constants $\mu_1$ and $\mu_2$ due to [$\sum_{k<l} (X_l - X_k)J_{kl}=0$]. Second, we use the Cauchy--Schwarz inequality in  \cref{eq:bounds-excess-1} to write 
\begin{align}
\label{eq:bounds-excess-2}
    \langle\!\langle y, x \rangle\!\rangle_- &\leq \sqrt{\langle\!\langle x, x \rangle\!\rangle_+}
    \big[\sqrt{ V_1} + \sqrt{ V_2}\big]\sqrt{\dot{\Pi}}
    \,,
\end{align}
where $\dot{\Pi} \equiv \sum_{k<l}J_{kl}^2/A_{kl}$ is the pseudo-entropy production~\cite{dechant2021improving,shiraishi2021optimal,van2022unified,dechant2022minimum} and where we introduced 
\begin{align}
     V_1 = \dot{\Pi}^{-1}\sum_{k<l} \frac{J_{kl}^2}{A_{kl}}(Y_{k} - \mu_1)^2\,,\quad  V_2 = \dot{\Pi}^{-1}\sum_{k<l} \frac{J_{kl}^2}{A_{kl}}(Y_{l} - \mu_2)^2\,.
\end{align}
Third, we choose $\mu_1$ and $\mu_2$ to minimize $ V_1$ and $ V_2$ as
\begin{align}
\label{eq:center}
   \frac{\partial V_i}{\partial \mu_i} = 0\,\rightarrow\,\mu_1 = \sum_{k<l}  M_{kl} Y_k\,,\,\, \mu_2 = \sum_{k<l} M_{kl} Y_l\,,
\end{align} 
with the weights $M_{kl}=\dot{\Pi}^{-1}J_{kl}^2/A_{kl}$ normalized as $\sum_{k<l}M_{kl}=1$. We note that $\mu_1$ and $\mu_2$ in \cref{eq:center} are the averages of the excess observables $\boldsymbol{Y}$ over the weights $\mathbb{M}$. Therefore, $V_1=\sum_{k<l}M_{kl}(Y_k - \mu_1)^2$ and $V_2=\sum_{k<l}M_{kl}(Y_l - \mu_2)^2$ are the weighted variances. Using the Bhatia--Davis bound first, followed by the Popoviciu inequalities for the variances $V_i$, we get
\begin{align}
\label{eq:bound-for-weighted-var}
    V_{i} \leq (\max_n Y_n - \mu_i)(\mu_i - \min_n Y_n) \leq \frac{(\Delta Y)^2}{4}\,,
\end{align}
where $i=1,2$ and $\Delta Y = \max_{n}Y_n - \min_{n}Y_n$. Similar bounds could be derived for $\boldsymbol{X}$ observables in terms of $\Delta X = \max_{n}X_n - \min_{n}X_n$. 
Using the second inequality in \cref{eq:bound-for-weighted-var} with \cref{eq:bounds-excess-1}, we arrive at the \textit{excess bound}
\begin{align}
\label{eq:bound-excess}
    |\langle\!\langle y, x \rangle\!\rangle_-| &\leq \min \big\{\Delta Y\langle\!\langle x, x \rangle\!\rangle_+^{1/2} \,, \Delta X \langle\!\langle y, y \rangle\!\rangle_+^{1/2} \big\} \dot{\Pi}^{1/2}\,,
\end{align}
We note that \cref{eq:bound-excess} can also be expressed in terms of the steady-state entropy production $\dot{\sigma}=\sum_{k<l}J_{kl}\ln(W_{kl}/W_{lk})$ since $\dot{\Pi}\leq \dot{\sigma}/2$.

\subsubsection{Example: Bounds for a simple molecular motor}

As an illustration of \cref{eq:bound-excess,eq:bound-geometric}, we consider the simple model for a molecular motor depicted in \cref{fig:fig-1}(b). All transitions are reversible and the one between states $1$ and $2$ is driven by chemical reservoirs generating a chemical potential difference $\Delta \mu$. 
\cref{fig:fig-1}(c) depicts the excess observables associated with the state observables $\boldsymbol{x} = (1,0,0)^\intercal$ and $\boldsymbol{y} = (0,1,0)^\intercal$ which are projectors on state $1$ and $2$, respectively. 
\cref{fig:fig-1}(d) depicts the AICov and its geometric and excess bound. 
At finite driving $\Delta \mu > 0$, the AICov measures the nonreciprocity of the cross-correlation between states $1$ and $2$. It vanishes at equilibrium ($\Delta \mu = 0$ and $\dot{\Pi} = 0$) where Onsager reciprocity is restored, but it also vanishes for large driving in this model. We also note that the bell-shaped behavior of the AICov as a function of $\Delta\mu$ is captured by the excess bound but not by the geometric bound.

\section{Continuum Limit: Diffusion}

Now we consider the results for SICov/AICov [\cref{sec:discr-exact-results}] in the continuum limit of the Markov jump process --- using diffusive scaling~\cite{falasco2023macroscopic} --- where the dynamics is described by a Fokker–Planck equation. 

\subsection{Setup}

We consider an occupation number vector in a $d$-dimensional lattice, $\boldsymbol{n}=(n_1, \dots, n_d)^\intercal\in \mathcal{Z}^d$, where $n_k$ takes values from $\{ 0,\dots, N-1\}$. A Markov jump process changes the occupation vector according to $\boldsymbol{n} \to \boldsymbol{n}+\Delta_{\rho}$, where $\rho$ denotes different mechanisms (e.g. reactions or reservoirs), and $\Delta_\rho = (\Delta_{1,\rho},\dots,\Delta_{d,\rho})^\intercal$ is a displacement vector. We also assume that for every forward transition $+\rho$ there exists a reversed transition $-\rho$, such that $\Delta_{\rho}=-\Delta_{-\rho}$.

The probability of the occupation vector $\boldsymbol{n}$ is denoted by $P_{\boldsymbol{n}}(t) \equiv P(t, \boldsymbol{n})$ and satisfies the master equation
\begin{align}
    d_t P(t, \boldsymbol{n}) &= \sum_{\boldsymbol{n}'}W_{\boldsymbol{n},\boldsymbol{n}'}P(\boldsymbol{n}',t)\,,\\
    &=\sum_\rho \Big[W_\rho(\boldsymbol{n}-\Delta_\rho)P(\boldsymbol{n}-\Delta_\rho\,,t) - W_\rho(\boldsymbol{n})P(\boldsymbol{n},t)\Big]\nonumber\,,
\end{align}
with the rate matrix $W_{\boldsymbol{n}\boldsymbol{n'}}=\sum_{\rho}W_\rho(\boldsymbol{n}')(\delta_{n',n-\Delta_{\rho}}-\delta_{n',n})$. Here, $W_\rho$ is the rate of transition $\rho\in\{+\rho,-\rho\}$. We assume that $W_{\boldsymbol{n}\boldsymbol{n'}}$ is irreducible such that the system relaxes to the unique steady state distribution satisfying to
\begin{align}
  \sum_{\boldsymbol{n}'}W_{\boldsymbol{n},\boldsymbol{n}'}P_\text{ss}(\boldsymbol{n}')=0\,.
\end{align}
We consider the limit $N\to\infty$ and introduce continuum variables $\boldsymbol{q} = \varepsilon \boldsymbol{n} \in \mathcal{R}^d$ with the lattice spacing $\varepsilon$ tending to zero $\varepsilon\to 0$. In this limit the probability, observables, and excess observables become functions of the coordinate vector $\boldsymbol{q}=(q_1,\dots,q_d)$: $p(t,\boldsymbol{q})$, $x(\boldsymbol{q})$, $X(\boldsymbol{q})$, respectively.

\subsection{Diffusive scaling}
\label{sec:CLimit}
We assume that the rates $W_\rho^{(\varepsilon)}$ depend on the small parameter $\varepsilon$ and can be decomposed as
\begin{align}
\label{eq:W-expansion}
    W_\rho^{(\varepsilon)} = \frac{\omega_\rho^\text{sym}}{\varepsilon^2} + \frac{\omega_\rho^\text{anti}}{\varepsilon}\,,
\end{align}
where $\omega_\rho^\text{sym} = \omega_{-\rho}^\text{sym}$ is the symmetric contribution and
$\omega_\rho^\text{anti} = - \omega_{-\rho}^\text{anti}$ is the antisymmetric one; they both do not depend on the small parameter $\varepsilon$; for details, see Eq.~(194) in \cite{falasco2023macroscopic}.
The discrete steady-state probability vector is normalized as $\sum_{n}P^{(\varepsilon)}(t, \boldsymbol{n})=1$, in the continuum limit we have the probability distribution $p(t, \boldsymbol{q}) \equiv P^{(\varepsilon)}(t, \boldsymbol{n})/\varepsilon^d$, such that $\int d\boldsymbol{q} p(t, \boldsymbol{q}) = 1$ where $\sum_{\boldsymbol{n}}\to\int d\boldsymbol{q}\varepsilon^{-d}$. The state based observables become functions of space $x(\boldsymbol{q})$ with mean $\langle x \rangle = \int d\boldsymbol{q} x(\boldsymbol{q})p^{\text{ss}}(\boldsymbol{q})$, where $p_\text{ss}(\boldsymbol{q}) \equiv P^{(\varepsilon)}_\text{ss}(\boldsymbol{n})/\varepsilon^d$ is the steady-state probability density.  

Next, we notice that the discrete displacement of the allowed jumps in continuous coordinates reads $\boldsymbol{q} \to \boldsymbol{q} + \varepsilon \Delta_\rho$. 
The first and second moments of these jumps define the drift $F_k$ and diffusion tensor in the limit $D_{kk'}$ as
\begin{subequations}
\label{eq:drift-D}
\begin{align}
\label{eq:drift}
   \mu_k &= \lim_{\varepsilon\to 0} \varepsilon \sum_{\rho}\Delta_{k\rho}W^{(\varepsilon)}_{\rho}=2\sum_{\rho>0}\Delta_{k\rho}\omega_\rho^\text{anti}\,,\\
\label{eq:diff}
   D_{kk'}&=\frac{1}{2}\lim_{\varepsilon\to 0} \varepsilon^2 \sum_{\rho}\Delta_{k\rho}\Delta_{k'\rho}W^{(\varepsilon)}_{\rho}=\sum_{\rho>0}\Delta_{k\rho}\Delta_{k'\rho}\omega^\text{sym}_{\rho}\,.
\end{align}
\end{subequations}
Similarly, using \cref{eq:W-expansion} we find the traffic along the transition $\Delta_{k\rho}$
\begin{align}
\label{eq:A-limit}
    A^{(\varepsilon)}_{\boldsymbol{n}+\Delta_{\rho}\boldsymbol{n}} &= (W^{(\varepsilon)}_{+\rho}+W^{(\varepsilon)}_{-\rho})(p_\text{ss}(\boldsymbol{q})+\mathcal{O}(\varepsilon))\varepsilon^d\nonumber\\
    &=2\omega^\text{sym}_\rho(\boldsymbol{q})p_\text{ss}(\boldsymbol{q})\varepsilon^{d-2}\,.
\end{align}
and the current,
\begin{align}
\label{eq:J-limit}
    \varepsilon^{-d}J^{(\varepsilon)}_{\boldsymbol{n}+\Delta_{\rho},\boldsymbol{n}} &= W^{(\varepsilon)}_{+\rho}p_\text{ss}(\boldsymbol{q})-W^{(\varepsilon)}_{-\rho}\Big(p_\text{ss}(\boldsymbol{q})+\varepsilon\sum_{k}\Delta_{k\rho}\frac{\partial p_\text{ss}(\boldsymbol{q})}{\partial x_k}\Big)\nonumber\\
    &\hspace{-0.5cm}=\frac{2\omega_\rho^\text{anti}(\boldsymbol{q})p_\text{ss}(\boldsymbol{q})}{\varepsilon} - \frac{\omega_\rho^\text{sym}(\boldsymbol{q})}{\varepsilon}\sum_{k}\Delta_{k\rho}\frac{\partial p_\text{ss}(\boldsymbol{q})}{\partial q_k}\,.
\end{align}

\subsection{Diffusive scaling of excess observables}
\label{sec:CLimit-X}

Probability density satisfies the Fokker--Planck equation
\begin{align}
    d_t p(t,\boldsymbol{q}) &= \mathcal{L} p(t,\boldsymbol{q})  \\
    &= -\sum_{k}\frac{\partial [\mu_k(\boldsymbol{q})  p(t,\boldsymbol{q})]}{\partial q_k}+\sum_{k,k'}\frac{\partial^2 [D_{kk'}(\boldsymbol{q})p(t, \boldsymbol{q})] }{\partial q_k\partial q_{k'}}\,,\nonumber
\end{align}
where $\mathcal{L}$ is the forward Fokker--Plank operator. The continuum current reads
\begin{align}
    j_k(t,\boldsymbol{q}) = - \mu_kp(t,\boldsymbol{q}) + \sum_{k'}D_{kk'}(\boldsymbol{q})\frac{\partial  p(t,\boldsymbol{q})}{\partial q_{k'}}\,.
\end{align}

The steady-state probability distribution $p_\text{ss}(\boldsymbol{q})$ satisfies the Fokker–Planck equation as
\begin{align}
    d_t p_\text{ss}(\boldsymbol{q}) &= \sum_{k}\frac{\partial j^\text{ss}_k}{\partial q_k} = 0 \;,\\
    j^\text{ss}_k(\boldsymbol{q}) &= - \mu_k(\boldsymbol{q}) p_\text{ss}(\boldsymbol{q}) + \sum_{k'}D_{kk'}(\boldsymbol{q})\frac{\partial  p_\text{ss}(\boldsymbol{q})}{\partial q_{k'}}\,,
\end{align}
where $j^\text{ss}_k(\boldsymbol{q})$ is the steady-state current, $D_{kl}(\boldsymbol{q})$ are the elements of the diffusion (postive semi-definite) matrix  and $\mu_k(\boldsymbol{q})$ is the drift field. The latter can be decomposed as 
\begin{align}
    \mu_k(\boldsymbol{q}) = - \sum_{k'}D_{kk'}(\boldsymbol{q})\frac{\partial \Phi(\boldsymbol{q})}{\partial q_{k'}} + f_k(\boldsymbol{q})
\end{align}
where $\Phi(\boldsymbol{q})$ is the state function and $\boldsymbol{f}$ is a force that could include non-potential terms (nonconservative forces). 

Now we find the continuum limit for the excess observable $X_{\boldsymbol{n}}$, which satisfies the Poisson equation $\sum_{\boldsymbol{n}}X_{\boldsymbol{n}}W^{(\varepsilon)}_{\boldsymbol{n}\boldsymbol{m}}=-(x_{\boldsymbol{m}} - \langle x \rangle)$. In the continuum limit, the left hand side of the Poisson equation reads
\begin{align}
   &\sum_{\boldsymbol{n}\neq \boldsymbol{m}}W^{(\varepsilon)}_{\boldsymbol{n}\boldsymbol{m}} (X_{\boldsymbol{n}} -X_{\boldsymbol{m}}) \to\sum_{\rho}W_{\rho}^{(\varepsilon)}(X(\boldsymbol{q}+\varepsilon\Delta_\rho) - X(\boldsymbol{q})) \nonumber\\
   &= \Big[\sum_k \mu_k \frac{\partial}{\partial q_k}  + \sum_{kk'}D_{kk'}\frac{\partial}{\partial q_k} \frac{\partial}{\partial q_{k'}}\Big] X(\boldsymbol{q}) = \mathcal{L}_\text{bwd} X(\boldsymbol{q})\,,
\end{align}
where we recognized the backward Kolmogorov operator $\mathcal{L}_\text{bwd}$. Thus $X(\boldsymbol{q})$ can be found solving
\begin{align}
\label{eq:Kolmogorov}
    \mathcal{L}_\text{bwd} X(\boldsymbol{q}) = - x(\boldsymbol{q}) + \langle x \rangle\,,
\end{align}
where $\langle x \rangle = \int d\boldsymbol{q}p_\text{ss}(\boldsymbol{q}) x(\boldsymbol{q})$. Thus, the continuum excess observable has the same meaning as the discrete \cref{eq:X-def},
\begin{align}
\label{eq:X-continuum}
    X(q) = \int_0^\infty dt (\langle x(t)|q\rangle - \langle x \rangle)\,.
\end{align}

\subsection{Diffusive scaling of integrated covariances}
\label{CLimit-results}
Here we derive the continuum limit counterparts of SICov and AICov \cref{eq:SICov-excess-1,eq:SICov-excess-2,eq:AICov-excess,eq:AICov-excess-2}.

\label{sec:continuum-exact-results}
\subsubsection{Continuum SICov}
Using \cref{eq:A-limit,eq:diff}, we calculate the continuum SIcov as
\begin{align}
\label{eq:SIcov-CL}
    \langle\!\langle x,x \rangle\!\rangle^+ &= \sum_{\boldsymbol{n}}\sum_{\rho>0} A^{\varepsilon}_{\boldsymbol{n}+\Delta_{\rho},\boldsymbol{n}}\big[X(\varepsilon\boldsymbol{n}+\varepsilon\Delta_\rho)- X(\varepsilon\boldsymbol{n})\big]^2\nonumber\\
    &=\int d\boldsymbol{q} \sum_{\boldsymbol{\rho}}W^{(\varepsilon)}_{\rho}p_\text{ss}(\boldsymbol{q})\varepsilon^2\sum_{k,k'}\Delta_{k\rho}\Delta_{k'\rho}\frac{\partial X(\boldsymbol{q})}{\partial q_{k}}\frac{\partial X(\boldsymbol{q})}{\partial q_{k'}}\nonumber\\
    &=2\sum_{k,k'}\int d\boldsymbol{q} D_{k,k'}(\boldsymbol{q})p_\text{ss}(\boldsymbol{q})\frac{\partial X(\boldsymbol{q})}{\partial q_{k}}\frac{\partial X(\boldsymbol{q})}{\partial q_{k'}}\,,
\end{align}
where $X(\boldsymbol{q})$ is the excess observable that satisfies the backward Kolmogorov \cref{eq:Kolmogorov}. 

\textit{Novelty remark---}We emphasize that \cref{eq:SIcov-CL} is known in the mathematical literature \cite{glynn1996liapounov,rey2015irreversible,duncan2016variance} as a nonequilibrium (nonreversible) extension of the Kipnis--Varadhan theorem \cite{kipnis1986central}. More recently, the same structure was rederived in stochastic thermodynamics~\cite{chun2026fluctuation}. In Ref.~\cite{chun2026fluctuation}, the authors used \cref{eq:SIcov-CL} to prove fluctuation–response relations, which play the role of the (Fokker--Planck) counterpart of the Markov jump process results \cite{aslyamov2025frr,ptaszynski2024frr,ptaszynski2025frr-mixed}.  
Similarly, \cref{eq:SICov-excess-1,eq:SICov-excess-2} are the microscopic (and discrete) counterparts of the continuum result \cref{eq:SIcov-CL}.

\subsubsection{Continuum AICov}
Using \cref{eq:J-limit,eq:drift-D} for AICov, we have 
\begin{align}
\label{eq:AICov-CLimit}
    &\langle\!\langle y,x \rangle\!\rangle^- = \sum_{\boldsymbol{n}}\sum_{\rho>0} J^{(\varepsilon)}_{\boldsymbol{n}+\Delta_{\rho},\boldsymbol{n}}\big[Y^{(\varepsilon)}_{\boldsymbol{n}+\Delta_{\rho}}X^{(\varepsilon)}_{\boldsymbol{n}}- X^{(\varepsilon)}_{\boldsymbol{n}+\Delta_{\rho}}Y^{(\varepsilon)}_{\boldsymbol{n}}\big]\nonumber\\
    &= \int d\boldsymbol{q}\sum_{\rho>0}
    \Big[ 2\omega_\rho^\text{anti}p_\text{ss} - \omega_\rho^\text{sym}\sum_{k'}\Delta_{k'\rho}\frac{\partial}{\partial q_{k'}}p_\text{ss}\Big]\nonumber\\
    &\times\sum_{k}\Delta_{\rho k}\Big[X \frac{\partial Y}{\partial q_k} - Y \frac{\partial X}{\partial q_k}\Big]\nonumber\\
    &=\int d\boldsymbol{q} \sum_{k} \Big[ \mu_k p_\text{ss} - \sum_{k'}D_{kk'}\frac{\partial}{\partial q_{k'}}p_\text{ss}\Big]\times\Big[X \frac{\partial Y}{\partial q_k} - Y \frac{\partial X}{\partial q_k}\Big] \nonumber\\
    &= \sum_{k}\int d\boldsymbol{q} j^\text{ss}_k(\boldsymbol{q})\Big[X(\boldsymbol{q})\frac{\partial Y(\boldsymbol{q})}{\partial q_k} - Y(\boldsymbol{q})\frac{\partial X(\boldsymbol{q})}{\partial q_k}\Big]\,,
\end{align}
which is a novel result. 

In the case of a general $d$, but simply connected region and reasonable boundary conditions (decay at infinity, or no net flux through the boundary), the steady-state current could be written as $j^\text{ss}_k=\sum_{\ell}\frac{\partial \mathcal{C}_{k\ell}}{\partial q_\ell}$ with $\mathcal{C}_{k\ell}=- \mathcal{C}_{\ell k}$, which allows us to rewrite \cref{eq:AICov-CLimit} as
\begin{align}
\label{eq:nonrecip-garad}
    \langle\!\langle y, x\rangle\!\rangle^- = 2\sum_{k,\ell}\int d\boldsymbol{q} \mathcal{C}_{k\ell}(\boldsymbol{q})\frac{\partial X(\boldsymbol{q})}{\partial q_k}\frac{\partial Y(\boldsymbol{q})}{\partial q_\ell}\,.
\end{align}
For $d=3$, the current is $\boldsymbol{j}^\text{ss}(\boldsymbol{q})=\nabla\times\boldsymbol{\mathcal{C}}$, where $\boldsymbol{\mathcal{C}}(\boldsymbol{q})$ is a vector field, then we have
\begin{align}
    \langle\!\langle y, x\rangle\!\rangle^- = 2\int d\boldsymbol{q}\, \boldsymbol{\mathcal{C}}\cdot (\nabla \boldsymbol{X} \times \nabla \boldsymbol{Y})\,.
\end{align}

\section{Application}
As an example of possible application we discuss the nonequilibrium speed up of self-averaging. 
\blue{The integrated autocorrelation time of observable $\boldsymbol{x}$ is defined as 
\begin{align}
\label{eq:autocorr-time-def}
    \tau_x \equiv \int_{-\infty}^\infty dt \frac{ C_{xx}(t)}{C_{xx}(0)} = \frac{\langle\!\langle x, x \rangle\!\rangle_+}{\langle \Delta x^2 \rangle}\,,
\end{align}
where $\langle \Delta x^2 \rangle = \sum_n \Delta x^2_n P_n^\text{ss}$. 
It is used to quantify the statistical uncertainty of the estimator for $\langle x \rangle$, since correlations between samples reduce the efficiency of averaging~\cite{sokal1997monte}.}
\blue{Strategies have been developed to accelerate self-averaging and reduce fluctuations by driving the system far from equilibrium, using MCMC \cite{neal2004improving,suwa2010markov,sun2010improving,bierkens2016non,coghi2021role} and Langevin equations~\cite{dechant2023thermodynamic,duncan2016variance}.
To meaningfully compare fluctuations of $\boldsymbol{x}$ out of equilibrium, $\langle\!\langle x,x\rangle\!\rangle_+$, and at equilibrium, $\langle\!\langle x,x\rangle\!\rangle_+^\text{eq}$, while keeping the same mean $\langle x \rangle = \langle x \rangle^\text{eq}$, the former are produced by a generic non-detailed balance rate matrix $\mathbb{W}$ with steady-state probability $\boldsymbol{P}^\text{ss}$, while the latter are generated by the detailed balance rate matrix $\mathbb{W}_\text{eq}$ with the same steady-state probability $\boldsymbol{P}^\text{ss} = \boldsymbol{P}^\text{eq}$.
Then it has been proven that $\mathbb{C}^+ \leq \mathbb{C}^+_\text{eq}$ \cite{sun2010improving}, which implies $\tau_x\leq\tau_x^\text{eq}$. We emphasize that $\tau_x$ is different from the relaxation time scale $\tau_\text{rlx}=1/|\text{Re}\lambda_2|$ for which it is known that $\tau_\text{rlx}\leq \tau_\text{rlx}^\text{eq}$~\cite{ichiki2013violation}.
}

\blue{To gain further insight, we wish to isolate the dissipative (energetic) contribution to the speed up. To do so, we request that the activity (kinetics) does not change: $\mathbb{A}=\mathbb{A}^\text{eq}$ in addition to $\boldsymbol{P}^\text{ss} = \boldsymbol{P}^\text{eq}$. 
This implies that $\mathbb{W}_\text{eq} = 1/2(\mathbb{W} + \mathbb{P}\cdot\mathbb{W}^\intercal \cdot\mathbb{P}^{-1})$ \cite{kolchinsky2024thermodynamic}.
We show in \cref{sec:speed-up} that this leads to the novel exact result:
\begin{align}
\label{eq:speed-up-exact}
\tau_x - \tau_x^\text{eq} = \frac{\boldsymbol{X} \cdot \mathbb{J} \cdot\boldsymbol{X}_\text{eq}}{\langle\Delta x^2\rangle} \leq 0\,,
\end{align}
where $\boldsymbol{X}_\text{eq}= - \boldsymbol{x}^\intercal\mathbb{W}_\text{eq}^D$ is the equilibrium excess observable and where we used $\langle \Delta x^2 \rangle^\text{eq} = \sum_n P_n^\text{eq} \Delta x_n^2 = \sum_n P_n^\text{ss} \Delta x_n^2 = \langle \Delta x^2 \rangle$.
}
\blue{The right hand side of \cref{eq:speed-up-exact} and \cref{eq:AICov-excess-2} have a similar structure.
Since the excess [\cref{eq:bound-excess}] and geometrical [\cref{eq:bound-geometric}] bounds can be applied to any $|\boldsymbol{X}^\intercal\mathbb{J}\boldsymbol{Y}|$, using \cref{eq:bound-geometric} for the right-hand side of \cref{eq:speed-up-exact}, we find
\begin{align}
\label{eq:bound-tau_x}
    0 \leq \frac{\tau_x^\text{eq} - \tau_x}{(\tau_x^\text{eq} + \tau_x)/2}\leq \mathcal{M}\,.
\end{align}
This can be used to bound the relative speed up of the self-averaging, $(\tau_x^\text{eq} - \tau_x)/\tau_x^\text{eq}$, using the cycle affinities as
\begin{align}
\label{eq:speed-up-bound}
0 \leq \frac{(\tau_x^\text{eq} - \tau_x)}{\tau_x^\text{eq}}\leq \frac{\mathcal{M}}{1+\mathcal{M}/2}\,.
\end{align}
For uni-cyclic networks as \cref{fig:fig-1}(b), the right-hand side of \cref{eq:speed-up-bound} is controlled by a single thermodynamic force. In \cref{fig:fig-1}(e), we illustrate the behavior of the exact \cref{eq:speed-up-exact} and the bound \cref{eq:speed-up-bound} as functions of the chemical potential difference driving the molecular motor. 
}

\begin{figure}
    \centering
    \includegraphics[width=\linewidth]{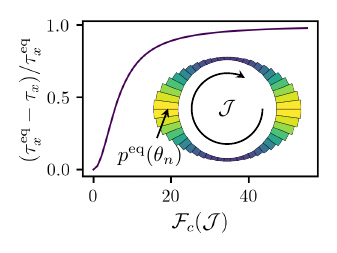}
    \caption{
Speed-up of self-averaging as a function of the cycle affinity generated by adding a clockwise cycle current $\mathcal{J}$ using the rotor model described in the main text.
Inset: Equilibrium distribution $p^{\mathrm{eq}}(\theta_n)$.
Simulations use $W^{\mathrm{eq}}_{n\pm1,n} = \exp[\beta V_0 \cos(2\theta_n)]$, $N=50$, $V_0 = 1$ and $\beta = 1$ with periodic boundary conditions $W^{\mathrm{eq}}_{1,N}=W^{\mathrm{eq}}_{N,1}$.
}
\label{fig:fig-2}
\end{figure}

\blue{This result also sheds light on the irreversible MCMC scheme. 
Indeed, the nonequilibrium generator 
\begin{align}
\label{eq:mcmc}
W_{nm}=W_{nm}^{\mathrm{eq}}+\frac{1}{2P_m^{\mathrm{eq}}}\sum_{\alpha}C_{nm}^{\alpha}\mathcal J_{\alpha}\,,
\end{align} 
produced from the equilibrium one by introducing controlled cycle currents $\mathcal J_{\alpha}$ satisfying $J_{nm}=\sum_{\alpha}C_{nm}^{\alpha} \mathcal J_{\alpha}< 2W_{nm}^{\mathrm{eq}}P_m^{\mathrm{eq}}$ (to prevent rates from becoming negative) with $C_{nm}^\alpha\in\{-1,0,1\}$ denoting the oriented cycle–edge incidence matrix depending on the edge orientation within cycle $\alpha$. \Cref{eq:mcmc} meets our requirements of preserving activity and the stationary distribution since $A^\text{eq}_{nm} = A_{nm}$ and $\sum_{m}W_{nm}^{\mathrm{eq}}P_m^\text{eq} = \sum_{m}W_{nm}P_m^\text{eq} = 0$ because of $C^\alpha_{nm}= - C^\alpha_{mn}$ and $\sum_{n}C^\alpha_{nm}=0$.
It can thus be used to speed up self-averaging. This procedure is equivalent to adding vorticity matrices in the irreversible MCMC algorithms \cite{bierkens2016non,sun2010improving}.} 

\blue{To illustrate \cref{eq:mcmc}, we consider a Markov jump process on a discretized ring with $N$ sites and coordinate $\theta_n = 2\pi n/N$. At equilibrium $P_n^{\mathrm{eq}}\!\propto\! \exp[-\beta V_0\cos(2\theta_n)]$, as represented in the inset of \cref{fig:fig-2}. 
This model may be thought of as describing the orientation of a colloidal or magnetic particle in a double-well aligning potential. 
Adding a clockwise cycle current $\mathcal{J}$ following \cref{eq:mcmc} produces a nonequilibrium process defined by the rates $W_{n\pm 1,n}=W_{n\pm 1,n}^{\mathrm{eq}}\pm \mathcal{J}/(2P_n^{\mathrm{eq}})$, which preserve the equilibrium distribution and activity.
The cycle current can vary as $0 \le \mathcal{J} < \min_n 2W_{n+1,n}^{\mathrm{eq}} P_n^{\mathrm{eq}}$ and produces a cycle affinity $\mathcal{F}_c(\mathcal{J})
= \sum_{n=1}^{N} \ln \frac{2W_{n+1,n}^{\mathrm{eq}}\pi_n^{\mathrm{eq}}+\mathcal{J}}{2W_{n,n+1}^{\mathrm{eq}} \pi_n^{\mathrm{eq}}-\mathcal{J}}$ (with $n+1$ taken modulo $N$). 
As seen in \cref{fig:fig-2}, this irreversible extension accelerates the self-averaging of observables $x=\cos\theta$, demonstrating how thermodynamic currents can enhance sampling efficiency without altering the target distribution.}

\textit{Continuum limit remark---}We notice that in the continuum limit of \cref{eq:speed-up-exact} reads
\begin{align}
\label{eq:speed-up-CLimit}
    \tau_x - \tau_x^\text{eq} = \frac{2}{\langle \Delta x^2 \rangle}\sum_{k,\ell}\int d\boldsymbol{q} \mathcal{C}_{k\ell}(\boldsymbol{q})\frac{\partial X(\boldsymbol{q})}{\partial q_k}\frac{\partial X^\text{eq}(\boldsymbol{q})}{\partial q_\ell}\,,
\end{align}
which considers two Fokker--Planck systems (one $\boldsymbol{f}=0$ and the other $\boldsymbol{f}\neq0$) which share the steady-state solution and diffusion matrix, but have different drift terms.

\section{Concluding remarks}

We have developed a unified formalism that connects SICov and AICov to excess observables \blue{which are operationally measureable quantities}. Our results provide exact expressions for fluctuations and nonreciprocity, clarifying their physical interpretations and relationships to thermodynamic quantities such as entropy production and activity. 
Our expressions in \cref{eq:SICov-excess-2,eq:AICov-excess-2} can also be seen as a numerically efficient way to compute the SICov and AICov using purely algebraic techniques and without resorting to time integration. 
\blue{In addition, we studied the continuum limit of our theory (diffusive scaling \cite{falasco2023macroscopic}) corresponding to the Fokker--Planck steady state. 
In this limit, the excess observable satisfies a Poisson equation involving the backward Kolmogorov operator, and \cref{eq:SICov-excess-2} becomes the microscopic discrete analogue of the nonequilibrium Kipnis--Varadhan-type variance representation \cite{duncan2016variance,glynn1996liapounov,kipnis1986central}; see \cref{eq:SIcov-CL}. For the antisymmetric expression \cref{eq:SICov-excess-2} and for the nonequilibrium speed-up~\cref{sec:speed-up}, we obtain new continuum formulas \cref{eq:AICov-CLimit,eq:speed-up-CLimit} that suggest connections to odd diffusion and nonreciprocal active matter~\cite{saha2020scalar, dinelli2023non}.
}

\begin{acknowledgments}
T.A. and M.E. acknowledge the financial support from, respectively, 
project ThermoElectroChem (C23/MS/18060819) from Fonds National de la Recherche-FNR, Luxembourg,  
project TheCirco (INTER/FNRS/20/15074473) funded by FRS-FNRS (Belgium) and FNR (Luxembourg).
The authors are grateful to Krzysztof Ptaszy\'{n}ski and Naruo Ohga for constructive comments on the manuscript. 
\end{acknowledgments}

\newpage

\appendix

\section{Derivation of \cref{eq:matC-results-1}}
\label{sec:derivation-matC}
Here we derive \cref{eq:matC-results-1} for SICov and ICov. 
The time integral of \cref{eq:corr-Markov} reads
\label{sec:matC}
\begin{align}
\label{eq:corr-int}
    \int_0^\infty C_{yx}(\tau) \,d\tau 
     &=  \Delta\boldsymbol{y}^\intercal \int_0^\infty e^{\tau \mathbb{W}}\big[\mathbb{1}  -\boldsymbol{P}^\text{ss}\boldsymbol{1}^\intercal \big]d\tau \,\mathbb{P}\Delta\boldsymbol{x} \nonumber\\ 
     &=  - \Delta\boldsymbol{y}^\intercal \mathbb{W}^D\mathbb{P}\Delta\boldsymbol{x} \,,
\end{align}
where we used $\boldsymbol{1}^\intercal\mathbb{P}\Delta \boldsymbol{x} = (\boldsymbol{P}^\text{ss})^\intercal \Delta\boldsymbol{x} = 0$ in the first identity and \cref{eq:Drazin-def} for $\mathbb{W}^D$. Using \cref{eq:corr-int} we calculate \cref{eq:Cov-def} as
\begin{align}
\label{eq:matC-deriviation-1}
    \langle\!\langle y, x\rangle\!\rangle_\pm &= \int_0^\infty (C_{yx}(\tau) \pm C_{xy}(\tau)) \,d\tau \\
    &= - \Delta\boldsymbol{y}^\intercal \big[\mathbb{W}^D\mathbb{P} \pm (\mathbb{W}^D\mathbb{P})^\intercal\big]\Delta\boldsymbol{x} = \Delta\boldsymbol{y}^\intercal \mathbb{C}^{\pm}\Delta\boldsymbol{x}\,,
\end{align}
with $\mathbb{C}^{\pm} \equiv - \big[\mathbb{W}^D\mathbb{P} \pm (\mathbb{W}^D\mathbb{P})^\intercal\big]$. 
Next, we derive \cref{eq:matC-results-1}. Using $\mathbb{W}^{D}\mathbb{P}\boldsymbol{1} = \mathbb{W}^{D}\boldsymbol{P}^\text{ss} = 0$ we write
\begin{align}
\label{eq:matC-deriviation-2}
    \mathbb{C}^{\pm} = -\big\{\mathbb{W}^D\mathbb{P}(\mathbb{1}-\boldsymbol{1}\boldsymbol{P}_\text{ss}^\intercal) \pm  \big[\mathbb{W}^D\mathbb{P}(\mathbb{1}-\boldsymbol{1}\boldsymbol{P}_\text{ss}^\intercal)\big]^\intercal\big\}\,.
\end{align}
From \cref{eq:Drazin-property} we notice that $\mathbb{1}-\boldsymbol{1}\boldsymbol{P}_\text{ss}^\intercal = (\mathbb{W}^D\mathbb{W})^\intercal=(\mathbb{W}\mathbb{W}^D)^\intercal$, so one can rewrite \cref{eq:matC-deriviation-1} as
\begin{align}
    \label{eq:matC-deriviation-3}
     \mathbb{C}^{\pm} &= -\big\{\mathbb{W}^D\mathbb{P}\mathbb{W}^\intercal(\mathbb{W}^D)^\intercal \pm  \mathbb{W}^D\mathbb{W}\mathbb{P}(\mathbb{W}^D)^\intercal\big\}\nonumber\\
     &=-\mathbb{W}^D(\mathbb{P}\mathbb{W}^\intercal \pm \mathbb{W}\mathbb{P})(\mathbb{W}^D)^\intercal\,,
\end{align}
which is \cref{eq:matC-results-1} in the main text.

\section{Derivation of \cref{eq:speed-up-exact}}
\label{sec:speed-up}
\blue{
Here we derive \cref{eq:speed-up-exact} using \cref{eq:matC-plus-result-1} as
\begin{align}
\label{eq:speed-up-deriviation-1}
    \langle\!\langle x,x\rangle\!\rangle &= - \Delta\boldsymbol{x}\cdot \mathbb{W}^{D}\cdot\mathbb{W}_\text{eq}\cdot\mathbb{W}_\text{eq}^D\cdot\mathbb{A}\cdot(\mathbb{W}^{D}\cdot\mathbb{W}_\text{eq}\cdot\mathbb{W}_\text{eq}^D)^\intercal\cdot\Delta\boldsymbol{x}\nonumber\\
    &=\Delta\boldsymbol{x}\cdot \mathbb{W}^{D}\cdot\mathbb{W}_\text{eq}\cdot\mathbb{C}_\text{eq}^+\cdot(\mathbb{W}^{D}\cdot\mathbb{W}_\text{eq})^\intercal\cdot\Delta\boldsymbol{x}\,,
\end{align}
where we used $\mathbb{W}^{D}\cdot\mathbb{W}_\text{eq}\cdot\mathbb{W}_\text{eq}^D = \mathbb{W}^{D} +\mathbb{W}^{D}\cdot\boldsymbol{P}^\text{ss}\boldsymbol{1}^\intercal = \mathbb{W}^{D}$ and $\mathbb{C}_\text{eq}^+ = - \mathbb{W}_\text{eq}^D\cdot\mathbb{A}\cdot(\mathbb{W}_\text{eq}^D)^\intercal$ due to $\boldsymbol{P}^\text{eq} = \boldsymbol{P}^\text{ss}$ and $\mathbb{A}^\text{eq} = \mathbb{A}^\text{ss}$.
Noticing that $\mathbb{W}_\text{eq}=\mathbb{W}-\tfrac{1}{2}\mathbb{J}\cdot\mathbb{P}^{-1}$ and $\Delta\boldsymbol{x}\cdot \mathbb{W}^{D}\cdot\mathbb{W}_\text{eq} = \Delta\boldsymbol{x} + \tfrac{1}{2}\boldsymbol{X}\cdot\mathbb{J}\cdot\mathbb{P}^{-1}$, we write \cref{eq:speed-up-deriviation-1} as
\begin{align}
    \label{eq:speed-up-deriviation-2}
    \langle\!\langle x,x\rangle\!\rangle - \langle\!\langle x,x\rangle\!\rangle_\text{eq} &= \frac{1}{2}\boldsymbol{X}\mathbb{J}\mathbb{P}^{-1}\mathbb{C}_\text{eq}^{+}\Delta\boldsymbol{x} + \frac{1}{2}\Delta\boldsymbol{x}(\boldsymbol{X}\mathbb{J}\mathbb{P}^{-1}\mathbb{C}_\text{eq}^{+})^{\intercal}\nonumber\\ 
    &+ \frac{1}{4}\boldsymbol{X}\mathbb{J}\mathbb{P}^{-1}\mathbb{C}_\text{eq}^{+}(\boldsymbol{X}\mathbb{J}\mathbb{P}^{-1})^{\intercal}\,.
\end{align}
Next, using $\mathbb{C}_\text{eq}^{+} = -2\mathbb{P}\cdot(\mathbb{W}_\text{eq}^D)^\intercal$ and $\boldsymbol{X}_\text{eq} = -\Delta \boldsymbol{x}\cdot \mathbb{W}^D_\text{eq}$, we notice that the first two terms on the right hand side of \cref{eq:speed-up-deriviation-2} sum to $2\boldsymbol{X}\cdot\mathbb{J}\cdot\boldsymbol{X}_\text{eq}$. 
Using $\mathbb{J} = 2(\mathbb{W} - \mathbb{W}_\text{eq})\mathbb{P}$ and $\mathbb{C}_\text{eq}^{+} = - 2\mathbb{W}_\text{eq}^D\mathbb{P}$ we write the third term as
\begin{align}
\label{eq:speed-up-deriviation-3}
    \frac{1}{4}\boldsymbol{X}\mathbb{J}\mathbb{P}^{-1}\mathbb{C}_\text{eq}^{+}(\boldsymbol{X}\mathbb{J}\mathbb{P}^{-1})^{\intercal} &= \boldsymbol{X}(\mathbb{W}-\mathbb{W}_\text{eq})\mathbb{W}_\text{eq}^{D}\mathbb{J}\boldsymbol{X} = \boldsymbol{X}_\text{eq}\mathbb{J}\boldsymbol{X}\nonumber\\
    &=-\boldsymbol{X}\mathbb{J}\boldsymbol{X}_\text{eq}\geq 0\,,
\end{align}
where we used \cref{eq:Poisson} for $\boldsymbol{X}\cdot\mathbb{W}\cdot\mathbb{W}_\text{eq}^D=-\Delta \boldsymbol{x}\cdot\mathbb{W}_\text{eq}^D = \boldsymbol{X}_\text{eq}$ and $\boldsymbol{X}\cdot\mathbb{W}_\text{eq}\cdot\mathbb{W}_\text{eq}^{D}\cdot\mathbb{J}\cdot\boldsymbol{X}=\boldsymbol{X}\cdot\mathbb{J}\cdot\boldsymbol{X} = 0$. \Cref{eq:speed-up-deriviation-3} is nonnegative because $\mathbb{C}_\text{eq}^{+}$ is positive semi-definite on the left hand side of \cref{eq:speed-up-deriviation-3}. Thus, we have 
\begin{align}
    \label{eq:speed-up-deriviation-4}
    \langle\!\langle x,x\rangle\!\rangle - \langle\!\langle x,x\rangle\!\rangle_\text{eq} &= \boldsymbol{X}\cdot\mathbb{J}\cdot\boldsymbol{X}_\text{eq}\leq 0\,,
\end{align}
which results in \cref{eq:speed-up-exact}. 
}

\bibliography{biblio}

@article{rey2015irreversible,
  title={Irreversible Langevin samplers and variance reduction: a large deviations approach},
  author={Rey-Bellet, Luc and Spiliopoulos, Konstantinos},
  journal={Nonlinearity},
  volume={28},
  number={7},
  pages={2081--2103},
  year={2015},
  publisher={IOP Publishing},
  doi = {10.1088/0951-7715/28/7/2081}
}

@article{chun2026fluctuation,
  title={Fluctuation-Response Theory for Nonequilibrium Langevin Dynamics},
  author={Chun, Hyun-Myung and Kwon, Euijoon and Park, Hyunggyu and Lee, Jae Sung},
  journal={arXiv preprint arXiv:2601.16387},
  year={2026},
  doi = {10.48550/arXiv.2601.16387}
}

@article{aslyamov2026macroscopic,
  title={Macroscopic fluctuation-response theory and its use for gene regulatory networks},
  author={Aslyamov, Timur and Ptaszy{\'n}ski, Krzysztof and Esposito, Massimiliano},
  journal={Physical Review Letters},
  volume={136},
  number={6},
  pages={067102},
  year={2026},
  publisher={APS},
  doi = {10.1103/2b12-3vrx}
}

@article{maes2020response,
  title={Response theory: a trajectory-based approach},
  author={Maes, Christian},
  journal={Frontiers in Physics},
  volume={8},
  pages={229},
  year={2020},
  publisher={Frontiers Media SA},
doi={10.3389/fphy.2020.00229}
}

@article{zheng2025nonlinear,
  title={Unified linear fluctuation-response theory arbitrarily far from equilibrium},
  author={Zheng, Jiming and Lu, Zhiyue},
  journal={Physical Review E},
  volume={112},
  number={6},
  pages={064103},
  year={2025},
  publisher={APS},
doi={10.1103/rgys-zxgf}
}

@article{yasuda2022time,
  title={Time-correlation functions for odd Langevin systems},
  author={Yasuda, Kento and Ishimoto, Kenta and Kobayashi, Akira and Lin, Li-Shing and Sou, Isamu and Hosaka, Yuto and Komura, Shigeyuki},
  journal={The Journal of Chemical Physics},
  volume={157},
  number={9},
  year={2022},
  publisher={AIP Publishing}
}

@article{coghi2021role,
  title={Role of current fluctuations in nonreversible samplers},
  author={Coghi, Francesco and Chetrite, Rapha{\"e}l and Touchette, Hugo},
  journal={Physical Review E},
  volume={103},
  number={6},
  pages={062142},
  year={2021},
  publisher={APS},
doi={10.1103/PhysRevE.103.062142}
}

@article{neal2004improving,
  title={Improving asymptotic variance of MCMC estimators: Non-reversible chains are better},
  author={Neal, Radford M},
  journal={arXiv preprint math/0407281},
  year={2004},
doi={10.48550/arXiv.math/0407281}
}

@article{bierkens2016non,
  title={Non-reversible metropolis-hastings},
  author={Bierkens, Joris},
  journal={Statistics and Computing},
  volume={26},
  number={6},
  pages={1213--1228},
  year={2016},
  publisher={Springer},
doi={10.1007/s11222-015-9598-x}
}

@article{sun2010improving,
  title={Improving the asymptotic performance of Markov chain Monte-Carlo by inserting vortices},
  author={Sun, Yi and Schmidhuber, J{\"u}rgen and Gomez, Faustino},
  journal={Advances in Neural Information Processing Systems},
  volume={23},
  year={2010},
url={https://proceedings.neurips.cc/paper_files/paper/2010/hash/819f46e52c25763a55cc642422644317-Abstract.html}
}

@article{ichiki2013violation,
  title={Violation of detailed balance accelerates relaxation},
  author={Ichiki, Akihisa and Ohzeki, Masayuki},
  journal={Physical Review E—Statistical, Nonlinear, and Soft Matter Physics},
  volume={88},
  number={2},
  pages={020101},
  year={2013},
  publisher={APS},
doi={10.1103/PhysRevE.88.020101}
}

@article{kolchinsky2024thermodynamic,
  title={Thermodynamic bound on spectral perturbations, with applications to oscillations and relaxation dynamics},
  author={Kolchinsky, Artemy and Ohga, Naruo and Ito, Sosuke},
  journal={Physical Review Research},
  volume={6},
  number={1},
  pages={013082},
  year={2024},
  publisher={APS},
doi={10.1103/PhysRevResearch.6.013082}
}

@article{suwa2010markov,
  title={Markov chain Monte Carlo method without detailed balance},
  author={Suwa, Hidemaro and Todo, Synge},
  journal={Physical review letters},
  volume={105},
  number={12},
  pages={120603},
  year={2010},
  publisher={APS},
doi={10.1103/PhysRevLett.105.120603}
}

@article{kipnis1986central,
  title={Central limit theorem for additive functionals of reversible Markov processes and applications to simple exclusions},
  author={Kipnis, Claude and Varadhan, SR Srinivasa},
  journal={Communications in Mathematical Physics},
  volume={104},
  number={1},
  pages={1--19},
  year={1986},
  publisher={Springer},
doi={10.1007/BF01210789}
}

@article{glynn1996liapounov,
  title={A Liapounov bound for solutions of the Poisson equation},
  author={Glynn, Peter W and Meyn, Sean P},
  journal={The Annals of Probability},
  pages={916--931},
  year={1996},
  publisher={JSTOR},
url={http://www.jstor.org/stable/2244957}
}

@article{duncan2016variance,
  title={Variance reduction using nonreversible Langevin samplers},
  author={Duncan, Andrew B and Lelievre, Tony and Pavliotis, Grigorios A},
  journal={Journal of statistical physics},
  volume={163},
  number={3},
  pages={457--491},
  year={2016},
  publisher={Springer},
doi={10.1007/s10955-016-1491-2}
}

@article{maes2008canonical,
  title={Canonical structure of dynamical fluctuations in mesoscopic nonequilibrium steady states},
  author={Maes, Christian and Neto{\v{c}}n{\`y}, Karel},
  journal={Europhysics Letters},
  volume={82},
  number={3},
  pages={30003},
  year={2008},
  publisher={IOP Publishing}
}

@article{garilli2025interrelation,
  title={Interrelation between precisions on integrated currents and on recurrence times in Markov jump processes},
  author={Garilli, Alberto and Frezzato, Diego},
  journal={Physical Review E},
  volume={112},
  number={4},
  pages={044141},
  year={2025},
  publisher={APS},
doi={doi.org/10.1103/27gn-7w5d}
}

@article{dechant2023thermodynamic,
  title={Thermodynamic bounds on correlation times},
  author={Dechant, Andreas and Garnier-Brun, Jerome and Sasa, Shin-ichi},
  journal={Physical Review Letters},
  volume={131},
  number={16},
  pages={167101},
  year={2023},
  publisher={APS},
doi={10.1103/PhysRevLett.131.167101}
}

@incollection{sokal1997monte,
  title={Monte Carlo methods in statistical mechanics: foundations and new algorithms},
  author={Sokal, Alan},
  booktitle={Functional integration: Basics and applications},
  pages={131--192},
  year={1997},
  publisher={Springer},
doi={10.1007/978-1-4899-0319-8_6}
}

@article{katayama2025diagrammatic,
  title={Diagrammatic expressions for steady-state distribution and static responses in population dynamics},
  author={Katayama, Koya and Nagayama, Ryuna and Ito, Sosuke},
  journal={arXiv preprint arXiv:2505.11296},
  year={2025},
doi={10.48550/arXiv.2505.11296}
}

@article{dinelli2023non,
  title={Non-reciprocity across scales in active mixtures},
  author={Dinelli, Alberto and O’Byrne, J{\'e}r{\'e}my and Curatolo, Agnese and Zhao, Yongfeng and Sollich, Peter and Tailleur, Julien},
  journal={Nature Communications},
  volume={14},
  number={1},
  pages={7035},
  year={2023},
  publisher={Nature Publishing Group UK London},
doi={10.1038/s41467-023-42713-5}
}

@article{saha2020scalar,
  title={Scalar active mixtures: The nonreciprocal Cahn-Hilliard model},
  author={Saha, Suropriya and Agudo-Canalejo, Jaime and Golestanian, Ramin},
  journal={Physical Review X},
  volume={10},
  number={4},
  pages={041009},
  year={2020},
  publisher={APS},
doi={10.1103/PhysRevX.10.041009}
}

@article{mitrophanov2025markov,
  title={Markov-Chain Perturbation and Approximation Bounds in Stochastic Biochemical Kinetics},
  author={Mitrophanov, Alexander Y},
  journal={Mathematics},
  year={2025},
  publisher={Multidisciplinary Digital Publishing Institute},
doi={10.3390/math13132059}
}

@article{tomita1974irreversible,
  title={Irreversible circulation of fluctuation},
  author={Tomita, Kazuhisa and Tomita, Hiroyuki},
  journal={Progress of theoretical physics},
  volume={51},
  number={6},
  pages={1731--1749},
  year={1974},
  publisher={Oxford University Press},
doi={10.1143/PTP.53.1546b}
}

@article{battle2016broken,
  title={Broken detailed balance at mesoscopic scales in active biological systems},
  author={Battle, Christopher and Broedersz, Chase P and Fakhri, Nikta and Geyer, Veikko F and Howard, Jonathon and Schmidt, Christoph F and MacKintosh, Fred C},
  journal={Science},
  volume={352},
  number={6285},
  pages={604--607},
  year={2016},
  publisher={American Association for the Advancement of Science},
doi={10.1126/science.aac8167}
}

@article{steinberg1986time,
  title={On the time reversal of noise signals},
  author={Steinberg, Izchak Z},
  journal={Biophysical journal},
  volume={50},
  number={1},
  pages={171--179},
  year={1986},
  publisher={Elsevier},
doi={10.1016/S0006-3495(86)83449-X}
}

@book{redner2001guide,
  title={A guide to first-passage processes},
  author={Redner, Sidney},
  year={2001},
  publisher={Cambridge university press}
}

@article{ptaszynski2025frr-mixed,
  title = {Nonequilibrium fluctuation-response relations for state-current correlations},
  author = {Ptaszy\ifmmode \acute{n}\else \'{n}\fi{}ski, Krzysztof and Aslyamov, Timur and Esposito, Massimiliano},
  journal = {Phys. Rev. E},
  volume = {113},
  issue = {2},
  pages = {024131},
  numpages = {11},
  year = {2026},
  month = {Feb},
  publisher = {American Physical Society},
  doi = {10.1103/4htr-dfc5}
}

@article{komatsu2009representation,
  title={Representation of nonequilibrium steady states in large mechanical systems},
  author={Komatsu, Teruhisa S and Nakagawa, Naoko and Sasa, Shin-Ichi and Tasaki, Hal},
  journal={Journal of Statistical Physics},
  volume={134},
  pages={401--423},
  year={2009},
  publisher={Springer},
doi={10.1007/s10955-009-9678-4}
}

@article{komatsu2008steady,
  title={Steady-state thermodynamics for heat conduction: microscopic derivation},
  author={Komatsu, Teruhisa S and Nakagawa, Naoko and Sasa, Shin-ichi and Tasaki, Hal},
  journal={Physical review letters},
  volume={100},
  number={23},
  pages={230602},
  year={2008},
  publisher={APS},
doi={10.1103/PhysRevLett.100.230602}
}

@article{khodabandehlou2024close,
  title={Close-to-equilibrium heat capacity},
  author={Khodabandehlou, Faezeh and Maes, Christian},
  journal={Journal of Physics A: Mathematical and Theoretical},
  volume={57},
  number={20},
  pages={205001},
  year={2024},
  publisher={IOP Publishing},
doi={10.1088/1751-8121/ad3ef2}
}

@article{bogers2025negative,
  title={Negative specific heats: where Clausius and Boltzmann entropies separate},
  author={Bogers, Lander and Khodabandehlou, Faezeh and Maes, Christian},
  journal={Physical Chemistry Chemical Physics},
  year={2025},
  publisher={Royal Society of Chemistry},
doi={10.1039/D5CP01269D}
}

@article{khodabandehlou2023nernst,
  title={A Nernst heat theorem for nonequilibrium jump processes},
  author={Khodabandehlou, Faezeh and Maes, Christian and Neto{\v{c}}n{\`y}, Karel},
  journal={The Journal of Chemical Physics},
  volume={158},
  number={20},
  year={2023},
  publisher={AIP Publishing},
doi={10.1063/5.0142694}
}

@article{khodabandehlou2024poisson,
  title={On the Poisson equation for nonreversible Markov jump processes},
  author={Khodabandehlou, Faezeh and Maes, Christian and Neto{\v{c}}n{\`y}, Karel},
  journal={Journal of Mathematical Physics},
  volume={65},
  number={4},
  year={2024},
  publisher={AIP Publishing},
doi={10.1063/5.0184909}
}

@article{ortner2024note,
  title={A Note on the Bias and Kemeny's Constant in Markov Reward Processes with an Application to Markov Chain Perturbation},
  author={Ortner, Ronald},
  journal={arXiv preprint arXiv:2408.04454},
  year={2024},
doi={10.48550/arXiv.2408.04454}
}

@article{pietzonka2016affinity,
  title={Affinity-and topology-dependent bound on current fluctuations},
  author={Pietzonka, Patrick and Barato, Andre C and Seifert, Udo},
  journal={Journal of Physics A: Mathematical and Theoretical},
  volume={49},
  number={34},
  pages={34LT01},
  year={2016},
  publisher={IOP Publishing},
doi={10.1088/1751-8113/49/34/34LT01}
}

@article{cho2000markov,
  title={Markov chain sensitivity measured by mean first passage times},
  author={Cho, Grace E and Meyer, Carl D},
  journal={Linear Algebra and its Applications},
  volume={316},
  number={1-3},
  pages={21--28},
  year={2000},
  publisher={Elsevier},
doi={10.1016/S0024-3795(99)00263-3}
}

@article{gu2024thermodynamic,
  title={Thermodynamic bounds on the asymmetry of cross-correlations with dynamical activity and entropy production},
  author={Gu, Jie},
  journal={Physical Review E},
  volume={109},
  number={4},
  pages={L042101},
  year={2024},
  publisher={APS},
doi={10.1103/PhysRevE.109.L042101}
}

@article{van2024dissipation,
  title={Dissipation, quantum coherence, and asymmetry of finite-time cross-correlations},
  author={Van Vu, Tan and Vo, Van Tuan and Saito, Keiji},
  journal={Physical Review Research},
  volume={6},
  number={1},
  pages={013273},
  year={2024},
  publisher={APS},
doi={10.1103/PhysRevResearch.6.013273}
}

@article{shiraishi2023entropy,
  title={Entropy production limits all fluctuation oscillations},
  author={Shiraishi, Naoto},
  journal={Physical Review E},
  volume={108},
  number={4},
  pages={L042103},
  year={2023},
  publisher={APS},
doi={10.1103/PhysRevE.108.L042103}
}

@article{qian2004fluorescence,
  title={Fluorescence correlation spectroscopy with high-order and dual-color correlation to probe nonequilibrium steady states},
  author={Qian, Hong and Elson, Elliot L},
  journal={Proceedings of the National Academy of Sciences},
  volume={101},
  number={9},
  pages={2828--2833},
  year={2004},
  publisher={National Academy of Sciences},
doi={10.1073/pnas.0305962101}
}

@article{liang2023thermodynamic,
  title={Thermodynamic bounds on time-reversal asymmetry},
  author={Liang, Shiling and Pigolotti, Simone},
  journal={Physical Review E},
  volume={108},
  number={6},
  pages={L062101},
  year={2023},
  publisher={APS},
doi={10.1103/PhysRevE.108.L062101}
}

@article{ptaszynski2024frr,
  title={Nonequilibrium Fluctuation-Response Relations for State Observables},
  author = {Ptaszy\ifmmode \acute{n}\else \'{n}\fi{}ski, Krzysztof and Aslyamov, Timur and Esposito, Massimiliano},
  journal = {Phys. Rev. E},
  volume = {113},
  issue = {2},
  pages = {024130},
  numpages = {9},
  year = {2026},
  month = {Feb},
  publisher = {American Physical Society},
  doi = {10.1103/r1qk-76gc}
}

@article{ptaszynski2024dissipation,
  title={Dissipation bounds precision of current response to kinetic perturbations},
  author={Ptaszy{\'n}ski, Krzysztof and Aslyamov, Timur and Esposito, Massimiliano},
  journal={Phys. Rev. Lett.},
  volume={133},
  number={22},
  pages={227101},
  year={2024},
  publisher={APS},
doi = {10.1103/PhysRevLett.133.227101}
}

@article{harvey2023universal,
  title={Universal energy-accuracy tradeoffs in nonequilibrium cellular sensing},
  author={Harvey, Sarah E and Lahiri, Subhaneil and Ganguli, Surya},
  journal={Physical Review E},
  volume={108},
  number={1},
  pages={014403},
  year={2023},
  publisher={APS},
doi={10.1103/PhysRevE.108.014403}
}

@article{kwon2024fluctuation,
  title={Fluctuation-response inequalities for kinetic and entropic perturbations},
  author={Kwon, Euijoon and Chun, Hyun-Myung and Park, Hyunggyu and Lee, Jae Sung},
  journal={arXiv preprint arXiv:2411.18108},
  year={2024},
doi = {10.48550/arXiv.2411.18108}
}

@book{stratonovich2012nonlinear,
  title={Nonlinear nonequilibrium thermodynamics I: linear and nonlinear fluctuation-dissipation theorems},
  author={Stratonovich, Rouslan L},
  volume={57},
  year={2012},
  publisher={Springer Science \& Business Media}
}

@article{shiraishi2023introduction,
  title={An Introduction to Stochastic Thermodynamics},
  author={Shiraishi, Naoto},
  journal={Fundamental Theories of Physics. Springer, Singapore},
  year={2023},
  publisher={Springer}
}

@book{kubo2012statistical,
  title={Statistical physics II: nonequilibrium statistical mechanics},
  author={Kubo, Ryogo and Toda, Morikazu and Hashitsume, Natsuki},
  volume={31},
  year={2012},
  publisher={Springer Science \& Business Media}
}

@article{dechant2019arxiv,
        title={Fluctuation-response inequality out of equilibrium}, 
      author={Andreas Dechant and Shin-ichi Sasa},
      year={2019},
  journal={arXiv preprint},
  year={2018},
 url={https://arxiv.org/abs/1804.08250}
}

@article{santos2020response,
  title={Response and sensitivity using Markov chains},
  author={Santos Guti{\'e}rrez, Manuel and Lucarini, Valerio},
  journal={Journal of Statistical Physics},
  volume={179},
  number={5},
  pages={1572--1593},
  year={2020},
  publisher={Springer},
doi={10.1007/s10955-020-02504-4}
}

@article{lucarini2016response,
  title={Response operators for Markov processes in a finite state space: radius of convergence and link to the response theory for Axiom A systems},
  author={Lucarini, Valerio},
  journal={Journal of Statistical Physics},
  volume={162},
  pages={312--333},
  year={2016},
  publisher={Springer},
doi={10.1007/s10955-015-1409-4}
}

@article{dechant2021improving,
  title={Improving thermodynamic bounds using correlations},
  author={Dechant, Andreas and Sasa, Shin-ichi},
  journal={Physical Review X},
  volume={11},
  number={4},
  pages={041061},
  year={2021},
  publisher={APS},
doi={10.1103/PhysRevX.11.041061}
}

@article{ohga2023thermodynamic,
  title={Thermodynamic bound on the asymmetry of cross-correlations},
  author={Ohga, Naruo and Ito, Sosuke and Kolchinsky, Artemy},
  journal={Physical Review Letters},
  volume={131},
  number={7},
  pages={077101},
  year={2023},
  publisher={APS},
doi={10.1103/PhysRevLett.131.077101}
}

@article{dechant2022minimum,
  title={Minimum entropy production, detailed balance and Wasserstein distance for continuous-time Markov processes},
  author={Dechant, Andreas},
  journal={Journal of Physics A: Mathematical and Theoretical},
  volume={55},
  number={9},
  pages={094001},
  year={2022},
  publisher={IOP Publishing},
doi={10.1088/1751-8121/ac4ac0}
}

@article{van2022unified,
  title={Unified thermodynamic--kinetic uncertainty relation},
  author={Van Vu, Tan and Hasegawa, Yoshihiko and others},
  journal={Journal of Physics A: Mathematical and Theoretical},
  volume={55},
  number={40},
  pages={405004},
  year={2022},
  publisher={IOP Publishing},
doi={10.1088/1751-8121/ac9099}
}

@article{shiraishi2021optimal,
  title={Optimal thermodynamic uncertainty relation in Markov jump processes},
  author={Shiraishi, Naoto},
  journal={Journal of Statistical Physics},
  volume={185},
  number={3},
  pages={19},
  year={2021},
  publisher={Springer},
doi={10.1007/s10955-021-02829-8}
}

@article{van2023thermodynamic,
  title={Thermodynamic unification of optimal transport: Thermodynamic uncertainty relation, minimum dissipation, and thermodynamic speed limits},
  author={Van Vu, Tan and Saito, Keiji},
  journal={Physical Review X},
  volume={13},
  number={1},
  pages={011013},
  year={2023},
  publisher={APS},
doi={10.1103/PhysRevX.13.011013}
}

@article{floyd2024limits,
  title={Limits on the computational expressivity of non-equilibrium biophysical processes},
  author={Floyd, Carlos and Dinner, Aaron R and Murugan, Arvind and Vaikuntanathan, Suriyanarayanan},
  journal={arXiv preprint},
  year={2024},
doi={10.48550/arXiv.2409.05827}
}

@article{frezzato2024steady,
  title={Steady-state probabilities for Markov jump processes in terms of powers of the transition rate matrix},
  author={Frezzato, Diego},
  journal={The Journal of Chemical Physics},
  volume={160},
  number={23},
pages={234111},
  year={2024},
  publisher={AIP Publishing},
doi={10.1063/5.0217202}
}

@article{di2018kinetic,
  title={Kinetic uncertainty relation},
  author={Di Terlizzi, Ivan and Baiesi, Marco},
  journal={Journal of Physics A: Mathematical and Theoretical},
  volume={52},
  number={2},
  pages={02LT03},
  year={2018},
  publisher={IOP Publishing},
  doi = {10.1088/1751-8121/aaee34}
}

@article{vu2020entropy,
  title={Entropy production estimation with optimal current},
  author={Van Vu, Tan and Vo, Van Tuan and Hasegawa, Yoshihiko},
  journal={Phys. Rev. E},
  volume={101},
  number={4},
  pages={042138},
  year={2020},
  publisher={APS},
doi = {10.1103/PhysRevE.101.042138}
}

@article{aslyamov2024general,
  title = {General Theory of Static Response for Markov Jump Processes},
  author = {Aslyamov, Timur and Esposito, Massimiliano},
  journal = {Phys. Rev. Lett.},
  volume = {133},
  issue = {10},
  pages = {107103},
  numpages = {7},
  year = {2024},
  month = {Sep},
  publisher = {American Physical Society},
  doi = {10.1103/PhysRevLett.133.107103},
  url = {https://link.aps.org/doi/10.1103/PhysRevLett.133.107103}
}

@article{gao2022thermodynamic,
  title={Thermodynamic constraints on the nonequilibrium response of one-dimensional diffusions},
  author={Gao, Qi and Chun, Hyun-Myung and Horowitz, Jordan M},
  journal={Phys. Rev. E},
  volume={105},
  number={1},
  pages={L012102},
  year={2022},
  publisher={APS},
doi={10.1103/PhysRevE.105.L012102}
}

@article{harunari2024mutual,
  title = {Mutual Linearity of Nonequilibrium Network Currents},
  author = {Harunari, Pedro E. and Dal Cengio, Sara and Lecomte, Vivien and Polettini, Matteo},
  journal = {Phys. Rev. Lett.},
  volume = {133},
  issue = {4},
  pages = {047401},
  numpages = {8},
  year = {2024},
  month = {Jul},
  publisher = {American Physical Society},
  doi = {10.1103/PhysRevLett.133.047401},
  url = {https://link.aps.org/doi/10.1103/PhysRevLett.133.047401}
}

@article{floyd2024learning,
  title={Learning to control non-equilibrium dynamics using local imperfect gradients},
  author={Floyd, Carlos and Dinner, Aaron R and Vaikuntanathan, Suriyanarayanan},
  journal={arXiv preprint},
  year={2024},
doi={10.48550/arXiv.2404.03798}
}

@article{gao2024thermodynamic,
  title={Thermodynamic constraints on kinetic perturbations of homogeneous driven diffusions},
  author={Gao, Qi and Chun, Hyun-Myung and Horowitz, Jordan M},
  journal={Europhys. Lett.},
  year={2024},
volume={146},
pages={31001},
doi={10.1209/0295-5075/ad40cd}
}

@article{gabriela2023topologically,
  title = {Topologically constrained fluctuations and thermodynamics regulate nonequilibrium response},
  author = {Fernandes Martins, Gabriela and Horowitz, Jordan M.},
  journal = {Phys. Rev. E},
  volume = {108},
  issue = {4},
  pages = {044113},
  numpages = {21},
  year = {2023},
  month = {Oct},
  publisher = {American Physical Society},
  doi = {10.1103/PhysRevE.108.044113},
  url = {https://link.aps.org/doi/10.1103/PhysRevE.108.044113}
}

@article{aslyamov2024nonequilibrium,
  title = {Nonequilibrium Response for Markov Jump Processes: Exact Results and Tight Bounds},
  author = {Aslyamov, Timur and Esposito, Massimiliano},
  journal = {Phys. Rev. Lett.},
  volume = {132},
  issue = {3},
  pages = {037101},
  numpages = {6},
  year = {2024},
  month = {Jan},
  publisher = {American Physical Society},
  doi = {10.1103/PhysRevLett.132.037101}
}

@article{aslyamov2025frr,
  title = {Nonequilibrium Fluctuation-Response Relations: From Identities to Bounds},
  author = {Aslyamov, Timur and Ptaszy\ifmmode \acute{n}\else \'{n}\fi{}ski, Krzysztof and Esposito, Massimiliano},
  journal = {Phys. Rev. Lett.},
  volume = {134},
  issue = {15},
  pages = {157101},
  numpages = {9},
  year = {2025},
  month = {Apr},
  publisher = {American Physical Society},
  doi = {10.1103/PhysRevLett.134.157101},
  url = {https://link.aps.org/doi/10.1103/PhysRevLett.134.157101}
}

@article{mallory2020kinetic,
  title={Kinetic control of stationary flux ratios for a wide range of biochemical processes},
  author={Mallory, Joel D and Kolomeisky, Anatoly B and Igoshin, Oleg A},
  journal={Proceedings of the National Academy of Sciences},
  volume={117},
  number={16},
  pages={8884--8889},
  year={2020},
  publisher={National Acad Sciences},
doi={10.1073/pnas.1920873117}
}

@article{agarwal1972fluctuation,
  title={Fluctuation-dissipation theorems for systems in non-thermal equilibrium and applications},
  author={Agarwal, Girish Saran},
  journal={Zeitschrift f{\"u}r Physik A Hadrons and nuclei},
  volume={252},
  number={1},
  pages={25--38},
  year={1972},
  publisher={Springer},
doi={10.1007/BF01391621}
}

@article{owen2020universal,
  title={Universal thermodynamic bounds on nonequilibrium response with biochemical applications},
  author={Owen, Jeremy A and Gingrich, Todd R and Horowitz, Jordan M},
  journal={Phys. Rev. X},
  volume={10},
  number={1},
  pages={011066},
  year={2020},
  publisher={APS},
doi={10.1103/PhysRevX.10.011066}
}

@article{maes2020frenesy,
  title={Frenesy: Time-symmetric dynamical activity in nonequilibria},
  author={Maes, Christian},
  journal={Physics Reports},
  volume={850},
  pages={1--33},
  year={2020},
  publisher={Elsevier},
doi={https://doi.org/10.1016/j.physrep.2020.01.002}
}

@article{prost2009generalized,
  title={Generalized fluctuation-dissipation theorem for steady-state systems},
  author={Prost, Jacques and Joanny, J-F and Parrondo, Juan MR},
  journal={Phys. Rev. Lett.},
  volume={103},
  number={9},
  pages={090601},
  year={2009},
  publisher={APS},
doi={10.1103/PhysRevLett.103.090601}
}

@article{seifert2010fluctuation,
  title={Fluctuation-dissipation theorem in nonequilibrium steady states},
  author={Seifert, Udo and Speck, Thomas},
  journal={Europhys. Lett.},
  volume={89},
  number={1},
  pages={10007},
  year={2010},
  publisher={IOP Publishing},
doi={10.1209/0295-5075/89/10007}
}

@article{baiesi2009fluctuations,
  title={Fluctuations and response of nonequilibrium states},
  author={Baiesi, Marco and Maes, Christian and Wynants, Bram},
  journal={Phys. Rev. Lett.},
  volume={103},
  number={1},
  pages={010602},
  year={2009},
  publisher={APS},
doi={10.1103/PhysRevLett.103.010602}
}

@article{altaner2016fluctuation,
  title={Fluctuation-dissipation relations far from equilibrium},
  author={Altaner, Bernhard and Polettini, Matteo and Esposito, Massimiliano},
  journal={Phys. Rev. Lett.},
  volume={117},
  number={18},
  pages={180601},
  year={2016},
  publisher={APS},
doi={10.1103/PhysRevLett.117.180601}
}

@article{kubo1966fluctuation,
  title={The fluctuation-dissipation theorem},
  author={Kubo, Rep},
  journal={Rep. Prog. Phys.},
  volume={29},
  number={1},
  pages={255},
  year={1966},
  publisher={IOP Publishing},
doi={10.1088/0034-4885/29/1/306}
}

@article{baiesi2013update,
  title={An update on the nonequilibrium linear response},
  author={Baiesi, Marco and Maes, Christian},
  journal={New Journal of Physics},
  volume={15},
  number={1},
  pages={013004},
  year={2013},
  publisher={IOP Publishing},
doi={10.1088/1367-2630/15/1/013004}
}

@article{forastiere2022linear,
  title={Linear stochastic thermodynamics},
  author={Forastiere, Danilo and Rao, Riccardo and Esposito, Massimiliano},
  journal={New Journal of Physics},
  volume={24},
  number={8},
  pages={083021},
  year={2022},
  publisher={IOP Publishing},
doi={10.1088/1367-2630/ac836b}
}

@article{rao2018conservation,
  title={Conservation laws shape dissipation},
  author={Rao, Riccardo and Esposito, Massimiliano},
  journal={New Journal of Physics},
  volume={20},
  number={2},
  pages={023007},
  year={2018},
  publisher={IOP Publishing},
doi={10.1088/1367-2630/aaa15f}
}

@article{owen2023size,
  title={Size limits the sensitivity of kinetic schemes},
  author={Owen, Jeremy A and Horowitz, Jordan M},
  journal={Nature Communications},
  volume={14},
  number={1},
  pages={1280},
  year={2023},
  publisher={Nature Publishing Group UK London},
doi={10.1038/s41467-023-36705-8}
}

@article{marconi2008fluctuation,
  title={Fluctuation--dissipation: response theory in statistical physics},
  author={Marconi, Umberto Marini Bettolo and Puglisi, Andrea and Rondoni, Lamberto and Vulpiani, Angelo},
  journal={Physics reports},
  volume={461},
  number={4-6},
  pages={111--195},
  year={2008},
  publisher={Elsevier},
doi={https://doi.org/10.1016/j.physrep.2008.02.002}
}

@article{barato2015thermodynamic,
  title={Thermodynamic uncertainty relation for biomolecular processes},
  author={Barato, Andre C and Seifert, Udo},
  journal={Phys. Rev. Lett.},
  volume={114},
  number={15},
  pages={158101},
  year={2015},
  publisher={APS},
doi={10.1103/PhysRevLett.114.158101}
}

@article{gingrich2016dissipation,
  title={Dissipation bounds all steady-state current fluctuations},
  author={Gingrich, Todd R and Horowitz, Jordan M and Perunov, Nikolay and England, Jeremy L},
  journal={Phys. Rev. Lett.},
  volume={116},
  number={12},
  pages={120601},
  year={2016},
  publisher={APS},
doi={10.1103/PhysRevLett.116.120601}
}

@article{falasco2019negative,
  title={Negative differential response in chemical reactions},
  author={Falasco, Gianmaria and Cossetto, Tommaso and Penocchio, Emanuele and Esposito, Massimiliano},
  journal={New Journal of Physics},
  volume={21},
  number={7},
  pages={073005},
  year={2019},
  publisher={IOP Publishing},
doi={10.1088/1367-2630/ab28be}
}

@article{falasco2023macroscopic,
  title = {Macroscopic stochastic thermodynamics},
  author = {Falasco, Gianmaria and Esposito, Massimiliano},
  journal = {Rev. Mod. Phys.},
  volume = {97},
  issue = {1},
  pages = {015002},
  numpages = {46},
  year = {2025},
  month = {Jan},
  publisher = {American Physical Society},
  doi = {10.1103/RevModPhys.97.015002},
  url = {https://link.aps.org/doi/10.1103/RevModPhys.97.015002}
}

@article{chun2023trade,
  title={Trade-offs between number fluctuations and response in nonequilibrium chemical reaction networks},
  author={Chun, Hyun-Myung and Horowitz, Jordan M},
  journal={The Journal of Chemical Physics},
  volume={158},
  number={17},
pages={174115},
  year={2023},
  publisher={AIP Publishing},
doi={https://doi.org/10.1063/5.0148662}
}

@article{dechant2020fluctuation,
  title={Fluctuation--response inequality out of equilibrium},
  author={Dechant, Andreas and Sasa, Shin-ichi},
  journal={Proceedings of the National Academy of Sciences},
  volume={117},
  number={12},
  pages={6430--6436},
  year={2020},
  publisher={National Acad Sciences},
doi={10.1073/pnas.1918386117}
}

@article{landi2023current,
  title = "{Current Fluctuations in Open Quantum Systems: Bridging the Gap Between Quantum Continuous Measurements and Full Counting Statistics}",
  author = {Landi, Gabriel T. and Kewming, Michael J. and Mitchison, Mark T. and Potts, Patrick P.},
  journal = {PRX Quantum},
  volume = {5},
  issue = {2},
  pages = {020201},
  numpages = {86},
  year = {2024},
  month = {Apr},
  publisher = {American Physical Society},
  doi = {10.1103/PRXQuantum.5.020201},
  url = {https://link.aps.org/doi/10.1103/PRXQuantum.5.020201}
}

@article{pietzonka2016universal,
  title = {Universal bounds on current fluctuations},
  author = {Pietzonka, Patrick and Barato, Andre C. and Seifert, Udo},
  journal = {Phys. Rev. E},
  volume = {93},
  issue = {5},
  pages = {052145},
  numpages = {16},
  year = {2016},
  month = {May},
  publisher = {American Physical Society},
  doi = {10.1103/PhysRevE.93.052145},
  url = {https://link.aps.org/doi/10.1103/PhysRevE.93.052145}
}

@article{pietzonka2017finite,
  title = {Finite-time generalization of the thermodynamic uncertainty relation},
  author = {Pietzonka, Patrick and Ritort, Felix and Seifert, Udo},
  journal = {Phys. Rev. E},
  volume = {96},
  issue = {1},
  pages = {012101},
  numpages = {6},
  year = {2017},
  month = {Jul},
  publisher = {American Physical Society},
  doi = {10.1103/PhysRevE.96.012101},
  url = {https://link.aps.org/doi/10.1103/PhysRevE.96.012101}
}

@article{horowitz2017proof,
  title = {Proof of the finite-time thermodynamic uncertainty relation for steady-state currents},
  author = {Horowitz, Jordan M. and Gingrich, Todd R.},
  journal = {Phys. Rev. E},
  volume = {96},
  issue = {2},
  pages = {020103},
  numpages = {3},
  year = {2017},
  month = {Aug},
  publisher = {American Physical Society},
  doi = {10.1103/PhysRevE.96.020103},
  url = {https://link.aps.org/doi/10.1103/PhysRevE.96.020103}
}

@article{falasco2020unifying,
  title={Unifying thermodynamic uncertainty relations},
  author={Falasco, Gianmaria and Esposito, Massimiliano and Delvenne, Jean-Charles},
  journal={New Journal of Physics},
  volume={22},
  number={5},
  pages={053046},
  year={2020},
  publisher={IOP Publishing},
doi={10.1088/1367-2630/ab8679}
}

@article{horowitz2020thermodynamic,
  title={Thermodynamic uncertainty relations constrain non-equilibrium fluctuations},
  author={Horowitz, Jordan M and Gingrich, Todd R},
  journal={Nature Physics},
  volume={16},
  number={1},
  pages={15--20},
  year={2020},
  publisher={Nature Publishing Group UK London},
doi={10.1038/s41567-019-0702-6}
}

@article{seifert2012stochastic,
  title={Stochastic thermodynamics, fluctuation theorems and molecular machines},
  author={Seifert, Udo},
  journal={Rep. Prog. Phys.},
  volume={75},
  number={12},
  pages={126001},
  year={2012},
  publisher={IOP Publishing},
doi={10.1088/0034-4885/75/12/126001}
}

@misc{crook2018drazin,
  author      = "Crooks, Gavin E.",
  title      = "On the {D}razin inverse of the rate matrix",
note="Technical note",
  year        = "2018",
url="https://threeplusone.com/pubs/drazin03/"
}

@article{berg1977physics,
  title={Physics of chemoreception},
  author={Berg, Howard C and Purcell, Edward M},
  journal={Biophysical J.},
  volume={20},
  number={2},
  pages={193--219},
  year={1977},
  publisher={Elsevier},
doi={10.1016/S0006-3495(77)85544-6}
}

@article{lapolla2020spectral,
  title = {Spectral theory of fluctuations in time-average statistical mechanics of reversible and driven systems},
  author = {Lapolla, Alessio and Hartich, David and Godec, A},
  journal = {Phys. Rev. Res.},
  volume = {2},
  issue = {4},
  pages = {043084},
  numpages = {17},
  year = {2020},
  month = {Oct},
  publisher = {American Physical Society},
  doi = {10.1103/PhysRevResearch.2.043084},
  url = {https://link.aps.org/doi/10.1103/PhysRevResearch.2.043084}
}

@article{lapolla2018unfolding,
  title={Unfolding tagged particle histories in single-file diffusion: exact single-and two-tag local times beyond large deviation theory},
  author={Lapolla, Alessio and Godec, Alja{\v{z}}},
  journal={New Journal of Physics},
  volume={20},
  number={11},
  pages={113021},
  year={2018},
  publisher={IOP Publishing},
doi={10.1088/1367-2630/aaea1b}
}

@article{lapolla2019manifestations,
  title={Manifestations of projection-induced memory: General theory and the tilted single file},
  author={Lapolla, Alessio and Godec, A},
  journal={Frontiers in Physics},
  volume={7},
  pages={182},
  year={2019},
  publisher={Frontiers Media SA},
doi={10.3389/fphy.2019.00182}
}

@article{ptaszynski2024critical,
  title = {{Critical heat current fluctuations in Curie-Weiss model in and out of equilibrium}},
  author = {Ptaszy\ifmmode \acute{n}\else \'{n}\fi{}ski, Krzysztof and Esposito, Massimiliano},
  journal = {Phys. Rev. E},
  volume = {111},
  issue = {3},
  pages = {034125},
  numpages = {15},
  year = {2025},
  month = {Mar},
  publisher = {American Physical Society},
  doi = {10.1103/PhysRevE.111.034125},
  url = {https://link.aps.org/doi/10.1103/PhysRevE.111.034125}
}

@article{jarzynski1997nonequilibrium,
  title = {Nonequilibrium Equality for Free Energy Differences},
  author = {Jarzynski, C.},
  journal = {Phys. Rev. Lett.},
  volume = {78},
  issue = {14},
  pages = {2690--2693},
  numpages = {0},
  year = {1997},
  month = {Apr},
  publisher = {American Physical Society},
  doi = {10.1103/PhysRevLett.78.2690},
  url = {https://link.aps.org/doi/10.1103/PhysRevLett.78.2690}
}

@article{crooks1999entropy,
  title = {Entropy production fluctuation theorem and the nonequilibrium work relation for free energy differences},
  author = {Crooks, Gavin E.},
  journal = {Phys. Rev. E},
  volume = {60},
  issue = {3},
  pages = {2721--2726},
  numpages = {0},
  year = {1999},
  month = {Sep},
  publisher = {American Physical Society},
  doi = {10.1103/PhysRevE.60.2721},
  url = {https://link.aps.org/doi/10.1103/PhysRevE.60.2721}
}

@article{esposito2009nonequilibrium,
  title = {Nonequilibrium fluctuations, fluctuation theorems, and counting statistics in quantum systems},
  author = {Esposito, Massimiliano and Harbola, Upendra and Mukamel, Shaul},
  journal = {Rev. Mod. Phys.},
  volume = {81},
  issue = {4},
  pages = {1665--1702},
  numpages = {0},
  year = {2009},
  month = {Dec},
  publisher = {American Physical Society},
  doi = {10.1103/RevModPhys.81.1665},
  url = {https://link.aps.org/doi/10.1103/RevModPhys.81.1665}
}

@article{lang2014thermodynamics,
  title = {Thermodynamics of Statistical Inference by Cells},
  author = {Lang, Alex H. and Fisher, Charles K. and Mora, Thierry and Mehta, Pankaj},
  journal = {Phys. Rev. Lett.},
  volume = {113},
  issue = {14},
  pages = {148103},
  numpages = {5},
  year = {2014},
  month = {Oct},
  publisher = {American Physical Society},
  doi = {10.1103/PhysRevLett.113.148103},
  url = {https://link.aps.org/doi/10.1103/PhysRevLett.113.148103}
}

@article{tesser2024out,
  title = {Out-of-Equilibrium Fluctuation-Dissipation Bounds},
  author = {Tesser, Ludovico and Splettstoesser, Janine},
  journal = {Phys. Rev. Lett.},
  volume = {132},
  issue = {18},
  pages = {186304},
  numpages = {7},
  year = {2024},
  month = {May},
  publisher = {American Physical Society},
  doi = {10.1103/PhysRevLett.132.186304},
  url = {https://link.aps.org/doi/10.1103/PhysRevLett.132.186304}
}

@article{khodabandehlou2025affine,
  title={Affine relationships between steady currents},
  author={Khodabandehlou, Faezeh and Maes, Christian and Neto{\v{c}}n{\`y}, Karel},
  journal={Journal of Physics A: Mathematical and Theoretical},
  volume={58},
  number={15},
  pages={155002},
  year={2025},
  publisher={IOP Publishing},
doi={10.1088/1751-8121/adc8ea}
}

@article{zheng2025spatial,
  title={Spatial Correlation Unifies Nonequilibrium Response Theory for Arbitrary Markov Jump Processes},
  author={Zheng, Jiming and Lu, Zhiyue},
  journal={arXiv preprint arXiv:2501.01050},
  year={2025},
doi = {10.48550/arXiv.2501.01050}
}

@article{bao2024nonequilibrium,
  title={Nonequilibrium Response Theory: From Precision Limits to Strong Perturbation},
  author={Bao, Ruicheng and Liang, Shiling},
  journal={arXiv preprint arXiv:2412.19602},
  year={2024},
doi = {10.48550/arXiv.2412.19602}
}

@misc{cengio2025mutual,
      title={Mutual Multilinearity of Nonequilibrium Network Currents}, 
      author={Sara Dal Cengio and Pedro E. Harunari and Vivien Lecomte and Matteo Polettini},
      year={2025},
      eprint={2502.04298},
      archivePrefix={arXiv},
      primaryClass={cond-mat.stat-mech}
}
\end{document}